%% file: main.tex
\documentclass[journal]{IEEEtran}

\usepackage[utf8]{inputenc}
\usepackage{times}
\usepackage[english]{babel}
\usepackage{amsmath}
\usepackage{amsfonts}
\usepackage{cleveref}
\usepackage{graphics}
\usepackage{comment}
\usepackage{tikz}
\usepackage{acro}
\usepackage{csquotes}
\usepackage{siunitx}
\usepackage{cite}

\usepackage{todonotes}

\input{macros_compat_paper.tex}

\input{own_macros.tex}

\newcommand{\pprimal}{\mat{P}}
\newcommand{\qprimal}{\mat{Q}}
\newcommand{\pdual}{\matdual{P}}
\newcommand{\qdual}{\matdual{Q}}

\newcommand{\widthscale}{1}

\graphicspath{{figures/}}

\begin{document}

\title{Magnetic and Combined Field Integral Equations Based on the Quasi-Helmholtz Projectors}

\author{Adrien~Merlini,~\IEEEmembership{Student~Member,~IEEE}\thanks{A.~Merlini is with the Department of Electronics and Telecommunications of Politecnico di Torino, Turin, Italy (e-mail: adrien.merlini@polito.it).},
Yves~Beghein\thanks{Y.~Beghein is with the Department of Information Technology (INTEC), Ghent University, 9502 Ghent, Belgium.}, %
Kristof~Cools\thanks{K.~Cools is with the Department of Electrical Engineering, Delft University of Technology, Delft, Netherlands},
Eric~Michielssen,~\IEEEmembership{Fellow,~IEEE}\thanks{E.~Michielssen is with the Department of Electrical Engineering and Computer Science, University of Michigan, Ann Arbor, MI 48109, USA (e-mail: emichiel@umich.edu).},\\
and Francesco~P.~Andriulli,~\IEEEmembership{Senior~Member,~IEEE}\thanks{F.~P.~Andriulli is with the Department of Electronics and Telecommunications of Politecnico di Torino, Turin, Italy (e-mail: francesco.andriulli@polito.it).}}

\maketitle

\begin{abstract}
Boundary integral equation methods for analyzing electromagnetic scattering phenomena typically suffer from several of the
following problems: (i) ill-conditioning when the frequency is low; (ii)
ill-conditioning when the discretization density is high; (iii) ill-conditioning when the structure contains global loops (which are computationally expensive to detect); (iv) incorrect solution at low frequencies due to current cancellations; (v) presence of spurious resonances.  In this paper,
quasi-Helmholtz projectors are leveraged to obtain \iac{MFIE} formulation
that is immune to drawbacks (i)-(iv). Moreover, when this
new \ac{MFIE} is combined with a regularized \ac{EFIE},  a new quasi-Helmholtz
projector \ac{CFIE} is obtained that also is immune to (v).
Numerical results corroborate the theory and show the
practical impact of the newly proposed formulations.
\end{abstract}

\acresetall

\section{Introduction}\label{sec:intro}

\IEEEPARstart{T}{ime-harmonic} scattering by \ac{PEC} objects oftentimes is modeled using frequency-domain boundary integral equations. Among them, electric and magnetic field integral equations (\acs{EFIE} and \acs{MFIE}) \cite{vanbladelElectromagneticFields2007} are the most popular.

Although the \ac{EFIE} is easily discretized using \ac{RWG} basis functions
\cite{raoElectromagneticScatteringSurfaces1982}, it suffers from
ill-conditioning when the frequency is low and/or the discretization density is
high. The \ac{MFIE}, on the other hand, remains well-conditioned in both regimes, provided  that a mixed discretization scheme is employed
\cite{coolsAccurateConformingMixed2011}.
In practice, however, it is not feasible to obtain
accurate results for the \ac{MFIE} at extremely low frequencies
without resorting to highly precise numerical quadrature methods. In
addition to the above issues, both the \ac{EFIE} and the \ac{MFIE} suffer from current
cancellations at low
frequencies \cite{yunhuazhangMagneticFieldIntegral2003,qianEnhancedAEFIEPerturbation2010,qianQuantitativeStudyLow2008}.

The \ac{EFIE}'s conditioning and current cancellation problems can be overcome by
using loop-star or loop-tree decompositions \cite{wiltonImprovingElectricField1981,
vecchiLoopstarDecompositionBasis1999, zhaoIntegralEquationSolution2000, leeLoopStarBasis2003,
eibertIterativesolverConvergenceLoopstar2004}. For multiply connected geometries, this
requires the detection of global loops, which is computationally expensive
\cite{wiltonImprovingStabilityElectric1981}. These techniques also fail to address the dense discretization breakdown phenomena \cite{andriulliSolvingEFIELow2010,
andriulliMultiplicativeCalderonPreconditioner2008} which causes the \ac{EFIE}'s condition number to grow
quadratically with the mesh refinement parameter. Worse still, loop-star techniques for combatting the \ac{EFIE}'s low-frequency conditioning problems further degrade the equations dense discretization behavior \cite{andriulliLoopStarLoopTreeDecompositions2012}.

Several formulations have been introduced to address these low-frequency issues
without the computational burden of global loop detection
\cite{qianAugmentedElectricField2008, zhuRigorousSolutionLowFrequency2011}. These solutions, however, do not
address the dense discretization ill-conditioning of the \ac{EFIE}.
Both issues can be concurrently tackled by leveraging hierarchical
quasi-Helmholtz decompositions \cite{vipianaMultiresolutionSystemRao2007,
andriulliHierarchicalBasesNonhierarchic2008, chenMultiresolutionCurvilinearRao2009, andriulliSolvingEFIELow2010}.
These decompositions also have been successfully coupled with other approaches such as \cald preconditioning
\cite{christiansenPreconditionerElectricField2002, contopanagosWellconditionedBoundaryIntegral2002, adamsPhysicalAnalyticalProperties2004,
darbasGeneralizedCombinedField2006, stephansonPreconditionedElectricField2009, andriulliMultiplicativeCalderonPreconditioner2008, suyanEFIEAnalysisLowFrequency2010,5654576,DOBBELAERE2015355,6934990} and
Debye-inspired schemes \cite{epsteinDebyeSourcesNumerical2010}. The price to be paid for this
dual stabilization is, once again, the need for global loop detection at very
low frequencies. In addition, several of the aforementioned techniques fail to
properly address low-frequency numerical cancellations occurring in the
solution vector \cite{yunhuazhangMagneticFieldIntegral2003, chewIntegralEquationMethods2008,
qianEnhancedAEFIEPerturbation2010, bogaertLowFrequencyScalingStandard2014}. Several of the above drawbacks have been successfully addressed by the  promising scheme in
\cite{vicoDecoupledPotentialIntegral2016}.
Alternative remedies to current cancellations include perturbation methods
\cite{yunhuazhangMagneticFieldIntegral2003, chewIntegralEquationMethods2008,
sunCalderonMultiplicativePreconditioned2013} and \cald regularization combined with 
loop star decompositions \cite{stephansonPreconditionedElectricField2009,
suyanEFIEAnalysisLowFrequency2010}.  Both families of solutions do, however, have shortcomings: the
former is only applicable at low frequencies and exhibits the same spectral
issues as the formulation it is applied to -- high refinement breakdown for
the \ac{EFIE} or global loop detection for the \ac{MFIE} and \cald
\ac{EFIE} -- while the latter also requires global loop detection and treatment of
the high refinement instability of the loop-star decomposition. It should also
be noted that some recent incarnations of augmented equations
are immune to several of the above
mentioned drawbacks, though they require the recovery of auxiliary quantities
\cite{chengAugmentedEFIENormally2015,dasModifiedSeparatedPotential2016}.

Recently, an electric type equation based on
quasi-Helmholtz projectors was proposed that is immune to all of
the aforementioned issues \cite{andriulliWellConditionedElectricField2013}.
A similar regularization has also been applied to
the time domain electric field integral equation
\cite{begheinDCStableLargeTime2015,begheinDCStableWellBalancedCalderon2015} and both the time domain and the
frequency domain PMCHWT equations
\cite{begheinHandlingLowfrequencyBreakdown2015,begheinRobustLowFrequency2015}.

In this paper, quasi-Helmholtz projectors are used to obtain
a new \ac{MFIE} that no longer requires interaction integrals to be computed using extremely accurate quadrature rules. Additionally, the solenoidal and nonsolenoidal
current components are scaled such that low frequency cancellations are
avoided. As a result, the formulation remains accurate down to extremely low
frequencies.  Scattering problems
involving \ac{PEC} objects can also be solved using the \ac{CFIE}, which is a linear combination of the \ac{EFIE} and the
\ac{MFIE}. This equation has the added benefit that it does not support spurious
resonances \cite{chewGedankenExperimentsUnderstand2007}. In this paper, the
new regularization method for the \ac{MFIE} is combined with that for the \ac{EFIE} presented in \cite{andriulliWellConditionedElectricField2013}. The resulting \ac{CFIE} is not only low-frequency stable but also immune to spurious resonances.
Preliminary results of this research
have previously been presented as conference contributions
\cite{andriulliMagneticTypeIntegral2014,andriulliWellconditionedCombinedField2013}.

This paper is organized as follows. To set notation, \Cref{sec:preliminaries} defines the standard \ac{EFIE} and \ac{MFIE} as well as their
discretizations and related quasi-Helmholtz current decompositions. In \Cref{sec:mfie}, a quasi-Helmholtz
decomposition is applied to a new symmetrized form of the \ac{MFIE}. The resulting
equation can be discretized accurately using standard numerical quadrature
methods, and can be scaled in frequency such that no low frequency
cancellations occur. In \Cref{sec:cfie}, this \ac{MFIE} is
combined with the regularized \ac{EFIE}
\cite{andriulliWellConditionedElectricField2013}
to obtain an extremely low frequency stable \ac{CFIE}.
\Cref{sec:numres} discusses numerical results that corroborate the
theory and conclusions are presented in \Cref{sec:conclusion}.

\section{Background and Notations}\label{sec:preliminaries}

The \ac{EFIE} and \ac{MFIE} operators $\Top_k$ and $\Kop_k$ are defined as
\begin{eqnarray}
\left(\Top_k\Jv\right)\left(\vr\right) & = & \left(\Top_{s,k}\Jv\right)\left(\vr\right)  +   \left(\Top_{h,k}\Jv\right)\left(\vr\right)\,,\\
\left(\Top_{s,k}\Jv\right)\left(\vr\right) & = &   \jm k \eta \n \times \int_\Gamma \frac{e^{-\jm k R}}{4 \uppi R} \Jv(\vrp) \dd s'\,,\\
\left(\Top_{h,k}\Jv\right)\left(\vr\right) & = &   - \frac{\eta}{\jm  k} \n \times \nabla \int_\Gamma \frac{e^{-\jm k R}}{4 \uppi R} \nabla' \cdot \Jv(\vrp) \dd s'\,,\\
\left(\Kop_k \Jv\right)\left(\vr\right) & = & - \n \times p.v. \int_\Gamma \nabla \times \frac{e^{-\jm k R}}{4 \uppi R} \Jv(\vr') \dd s'\,,
\end{eqnarray}
where $R=\|\vr-\vr'\|$, $\Gamma$ is the boundary of a closed domain $\Omega \subset \R^3$ and $\n$ is its exterior normal vector. Furthermore, given the angular frequency $\omega$, $k = \omega \sqrt{\mu \epsilon}$ and $\eta = \sqrt{\mu / \epsilon}$; here $\epsilon$ and $\mu$ the permittivity and permeability of vacuum, respectively.
If $\Omega$ is perfectly conducting, it supports an electric current $\Jv(\vr)$ satisfying both the \ac{EFIE}
\begin{equation}
    \left(\Top_k \Jv\right)(\vr) = \n \times \Ev^i(\vr)  \label{eq:contefie}
\end{equation}
and the \ac{MFIE}
\begin{equation}
    \left(\left(\frac{\Iop}{2} + \Kop_k\right)\Jv\right)(\vr) = \n \times \Hv^i(\vr) \label{eq:contmfie}
\end{equation}
for all $\vr \in \Gamma$; where $\Ev^i$ and $\Hv^i$ denote the impinging electric and magnetic fields,
respectively.
To numerically solve these equations via a Galerkin procedure,
 $\Jv(\vr)$ is expanded into RWG basis functions
$\{\veg f_j(\vr)\}$ \cite{raoElectromagneticScatteringSurfaces1982} as
\begin{equation}
    \Jv(\vr) \approx \sum_{j=1}^{N_e} \left[\Jvec\right]_j \veg{f}_j(\vr)\,,
\end{equation}
where $N_e$ is the number of edges of the mesh.
Following \cite{andriulliWellConditionedElectricField2013}, the RWG functions are normalized such that the integrated flux through their defining edges equals one.
%
\begin{figure}
    \centering
    \begin{tikzpicture}
        \tikzstyle{vertex}=[circle,thick,draw=black!75,fill=red,minimum size=2mm]

        \node []       (m)                                       {};
        \node []       (ml) [left of=m]                   {$c_n^+$};
        \node []       (mr) [right of=m]                  {$c_n^-$};
        \node [vertex] (a) [label=above:$\veg{v}_n^+$, above of=m]     {};
        \node [vertex] (c) [label=below:$\veg{v}_n^-$, below of=m]     {};
        \node [vertex] (b) [label=left:$\vr_n^+$, left of=ml]      {};
        \node [vertex] (d) [label=right:$\vr_n^-$, right of=mr]    {};

        \draw (b) -- (a);
        \draw (a) -- (d);
        \draw (b) -- (c);
        \draw (c) -- (d);
        \draw[ultra thick, -latex, label=$\veg{e}_n$] (c) -- (a) node [midway, label=left:$\veg{e}_n$] {};
    \end{tikzpicture}
    \caption{Notations used for the definition of an RWG basis function; $\veg{e}_n$ denotes the defining inner edge that links vertices $\veg{v}_n^+$ and $\veg{v}_n^-$ and $c_n^+$ and $c_n^-$ the two triangles connected to this edge which are completed by the vertices $\vr_n^+$ and $\vr_n^-$, respectively.}
    \label{fig:RWGconv}
\end{figure}
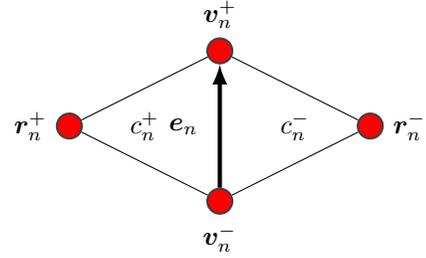
Next, the \ac{EFIE} \eqref{eq:contefie} is tested with rotated RWG functions $\{\n \times \veg{f}_i(\vr)\}$, while the \ac{MFIE} \eqref{eq:contmfie} is tested with rotated \ac{BC} functions \cite{buffaDualFiniteElement2007} $\{\n \times \veg{g}_i(\vr)\}$. The \ac{BC} functions $\{\veg g_j \}$ are divergence-conforming functions defined on the barycentric refinement of the mesh. In addition, they are quasi curl-conforming in the sense that the mixed Gram matrix between curl-conforming rotated \ac{BC} functions and \ac{RWG} functions is well conditioned.
For an explicit definition of these functions the reader is referred to \cite{andriulliMultiplicativeCalderonPreconditioner2008,buffaDualFiniteElement2007}.
Overall, the testing procedure results in the following matrix equations:
\begin{eqnarray}
    \mat{T} \Jvec & = & \vec{v}_e \label{EFIE},\\
    \left( \frac{ \mat{G}^\T }{2} + \mat{K}_{k} \right) \Jvec & = & \vec{v}_h \label{MFIE},
\end{eqnarray}
where
\begin{eqnarray}
\left[\mat{T}\right]_{ij} & = & \left(\n \times \veg{f}_i, \Top_k \veg{f}_j\right),\\
\left[\mat{T}_s\right]_{ij} & = & \left(\n \times \veg{f}_i, \Top_{s,k} \veg{f}_j\right),\\
\left[\mat{T}_h\right]_{ij} & = & \left(\n \times \veg{f}_i, \Top_{h,k} \veg{f}_j\right),\\
\left[\mat{K}_k\right]_{ij} & = & \left(\n \times \veg{g}_i, \Kop_k \veg{f}_j\right),\\
\left[\mat{G}\right]_{ij} & = & \left(\veg{f}_i, \n \times \veg{g}_j \right),\\
\left[\vec{v}_e\right]_i & = & \left(\n \times \veg{f}_i, \n \times \Ev^i\right),\\
\left[\vec{v}_h\right]_i & = & \left(\n \times \veg{g}_i, \n \times \Hv^i\right),
\end{eqnarray}
with $\left(\veg{a}, \veg{b}\right) = \int_\Gamma \veg{a}(\vr) \cdot \veg{b}(\vr) \dd s$. In addition we denote by $\matdual{T}$, $\matdual{T}_s$ and $\matdual{T}_h$ the \ac{BC}-expanded and tested counterparts of the discretized operators $\mat{T}$, $\mat{T}_s$ and $\mat{T}_h$ computed with the complex wavenumber $-\jm k$.

The solutions of \eqref{EFIE} and \eqref{MFIE} can be expressed as linear combinations of divergence free (loop and harmonic functions) and of non-divergence free (star functions) contributions via a quasi-Helmholtz decomposition
\begin{equation}\label{eq:firstIdec}
  \Jvec = \mat{\Lambda}\vec{l}+\mat{\Sigma}\vec{s}+\mat{H}\vec{h}
\end{equation}
where the first two matrices $\mat{\Lambda} \in \R^{N_e \times N_v}$ and
$\mat{\Sigma} \in \R^{N_e \times N_f}$ represent mappings from the
RWG subspace to the local loop and star subspaces, respectively. Here, $N_v$ and $N_f$
are the number of vertices and facets of the mesh, respectively \cite{wiltonImprovingStabilityElectric1981,coolsNullspacesMFIECalderon2009}.
These two mappings can be defined using only the connectivity information of the discretized geometry as
\begin{align}
  \mat{\Lambda}_{ij} &=
  \begin{cases}
     1 & \text{if node } j \text{ equals } \veg v_i^+\\
    -1 & \text{if node } j \text{ equals } \veg v_i^-\\
     0 & \text{otherwise}
  \end{cases} \label{eq:loopmat}\\
\intertext{and}
  \mat{\Sigma}_{ij} &=
  \begin{cases}
     1 & \text{if the cell } j \text{ equals } c_i^+\\
    -1 & \text{if the cell } j \text{ equals } c_i^-\\
     0 & \text{otherwise}\,,
  \end{cases} \label{eq:starmat}
\end{align}
where vertices $\veg v_i^-$ and $\veg v_i^+$ define the oriented edge characterizing RWG function $i$, and $c_i^-$ and $c_i^+$ denote the corresponding cells (\Cref{fig:RWGconv}).
The matrix $\mat{H}$ represents the mapping from the RWG space to the
quasi-harmonic or \textit{global loop} space composed of $2 N_h$ functions, where $N_h$ is the number of handles in the structure. For a complete description of this mapping and the associated harmonic functions, the reader is referred to \cite{wiltonImprovingStabilityElectric1981} and \cite{coolsNullspacesMFIECalderon2009}.

A few properties of these matrices are recalled next to facilitate
further developments. For the sake of simplicity we restrict ourselves to the
case of a geometry with a single closed connected component. All derivations below can be extended to arbitrary geometries using the relations in \cite{wiltonTopologicalConsiderationSurface1983}. Given this assumption, $\mat{\Lambda}$ has a
null-space spanned by the all-one vector $\veg{1}^{\mat \Lambda} \in
\R^{N_v}$, i.e.
\begin{equation}
  \mat{\Lambda}\veg{1}^{\mat \Lambda}=\vegO.
\end{equation}
Similarly, linear dependency of the star functions cause $\mat{\Sigma}$ to exhibit a one-dimensional null-space spanned by the all-one vector $\veg{1}^{\mat \Sigma} \in \R^{N_f}$, i.e.
\begin{equation}
  \mat{\Sigma}\veg{1}^{\mat \Sigma}=\vegO.
\end{equation}
Finally, it is trivial to show that the loop and star subspaces are orthogonal, i.e.
\begin{equation}\label{eq:orthloopstar}
  \mat{\Sigma}^{\T}\mat{\Lambda}=\matO.
\end{equation}

As $\matL$ and $\matS$ are ill-conditioned and because of the high computational cost of detecting global loops required to build $\mat H$, it is convenient to leverage
the quasi-Helmholtz projectors introduced in
\cite{andriulliWellConditionedElectricField2013} to obtain a quasi-Helmholtz
decomposition of the \ac{EFIE} and \ac{MFIE} operators. The projectors are defined as
\begin{align}
    \pstar &= \mat{\Sigma} \pseudoinv{\left( \mat{\Sigma}^\T \mat{\Sigma} \right)} \mat{\Sigma}^\T\,,\\
    \ploop &= \matId - \pstar\,,
\end{align}
where $\pseudoinv{}$ denotes the Moore-Penrose pseudo-inverse and $\matId$ is the identity. Any RWG
expansion coefficient vector can then be decomposed as
\begin{equation}
    \Jvec = \left(\ploop \Jvec\right) + \left(\pstar \Jvec\right)
\end{equation}
where $\ploop \Jvec$ and $\pstar \Jvec$ contain the RWG expansions of the solenoidal (loop) and non-solenoidal (star) components of the current, respectively.
These operators are self-adjoint and
also can be used to decompose the RWG testing space. Similarly, the dual
projectors $\ppstar$ and $\pploop$, defined as
\begin{align}
    \ppstar &= \mat{\Lambda} \pseudoinv{\left( \transpose{\mat{\Lambda}} \mat{\Lambda} \right)} \transpose{\mat{\Lambda}},\\
    \pploop &= \matId - \ppstar,
\end{align}
decompose any linear combination of \ac{BC} (basis or testing) functions into a
non-solenoidal and solenoidal part, respectively. It should be noted that
construction of these projectors does not require the detection of global loops,
and that $\pseudoinv{\left( \transpose{\mat{\Sigma}} \mat{\Sigma} \right)}$ can
be efficiently computed using multigrid preconditioners
\cite{andriulliWellConditionedElectricField2013, napovAlgebraicMultigridMethod2012}.


\section{Regularizing the \ac{MFIE} at Extremely Low Frequencies}\label{sec:mfie}

\subsection{Low Frequency Behaviour of the \ac{MFIE}}

The standard RWG discretization of the \ac{MFIE} fails to provide accurate results
at low frequencies due to the unphysical scaling of the loop and star (or
tree) components of the current \cite{yunhuazhangMagneticFieldIntegral2003}. It was shown in
\cite{coolsImprovingMFIEAccuracy2009,coolsAccurateConformingMixed2011} that the mixed discretization of
the \ac{MFIE} (in which \ac{BC} or CW functions \cite{chenElectromagneticScatteringThreedimensional1990} are used as testing functions) improves
accuracy. In particular, the loop and star components of the current
obtained from this formulation scale physically \cite{bogaertLowFrequencyScalingStandard2014}. This
result also holds true for multiply connected geometries \cite{bogaertLowFrequencyScaling2011}.

The mixed \ac{MFIE} formulation still suffers from three problems. First, the
physical scaling of the current can only be retrieved when interaction integrals are
computed to high accuracy \cite{bogaertLowFrequencyScalingStandard2014}.
Second, the nonsolenoidal current
component scales as $\order{\omega}$ whereas the solenoidal component is
of $\order{1}$. As a result, at very low frequencies and  when using finite precision, both components should be stored in
different arrays to prevent the nonsolenoidal component
from losing accuracy or even being cancelled out
\cite{yunhuazhangMagneticFieldIntegral2003,chewIntegralEquationMethods2008,qianEnhancedAEFIEPerturbation2010,bogaertLowFrequencyStability2011,bogaertLowFrequencyScaling2011}.
Third, the static \ac{MFIE} (at $\omega = 0$) has
a null space when applied to multiply connected geometries. It follows that the
discretized \ac{MFIE} has $N_h$ singular values that scale as $\order{\omega^2}$ \cite{coolsNullspacesMFIECalderon2009}. Any
accurate discretization of the \ac{MFIE} operator must preserve this null-space. Standard
RWG discretizations of the \ac{MFIE} operators are not capable of correctly
modelling this null space \cite{andriulliMagneticTypeIntegral2014}. The mixed \ac{MFIE}, on the
other hand, correctly models this-null space in infinite precision.
However, after discretization, the null-space associated singular values will not be more accurate than the precision of the quadrature rule.


\subsection{A Robust \ac{MFIE} Formulation}
To address the above described \ac{MFIE} deficiencies we propose the following symmetrized MFIE:
\begin{equation}
    \left( \frac{\Iop}{2} - \Kop_{-\jm k} \right) \left(\frac{\Iop}{2} + \Kop_k \right)  \left(\Jv\right) = \left( \frac{\Iop}{2} - \Kop_{-\jm k} \right)  \left( \n_r \times \Hv^i \right). \label{eq:symmfie}
\end{equation}
This equation is the magnetic field counterpart of the (localized) \cald preconditioned electric operator in
\cite{andriulliWellConditionedElectricField2013}.
We propose to discretize \eqref{eq:symmfie} as
\begin{eqnarray}
& & \matdual{M}^\T \left( \frac{ \mat{G}^\T }{2} - \mat{K}_{-\jm k} \right) \left( \mat{G}^\T \right)^{-1} \left( \frac{ \mat{G}^\T }{2} + \mat{K}_{k} \right) \mat{M} \vec{i} \nonumber \\
&=& \mat{O}\vec{i}  =  \matdual{M}^\T \left( \frac{ \mat{G}^\T }{2} - \mat{K}_{-\jm k} \right) \left( \mat{G}^\T \right)^{-1} \vec{v}_h \label{eq:discretizedmfie}
\end{eqnarray}
where
\begin{eqnarray}
    \mat{M} & = & \ploop \frac{1}{\alpha} + \jm \pstar \alpha\,,\\
    \matdual{M} & = & \pploop \frac{1}\alpha + \jm \ppstar \alpha\,,
\end{eqnarray}
and $\mat{M} \vec{i} = \Jvec$.

The coefficient $\alpha$ allows for re-scaling of the loop and star components of
the solution $\vec{i}$ of \eqref{eq:discretizedmfie} to prevent
numerical cancellations. Because $\pstar+\ploop = \ppstar+\pploop = \matId$,
operator $\mat{O}$ in \eqref{eq:discretizedmfie} can be decomposed as
\begin{multline} \label{eq:decomposed_O}
    \mat{O}=(\ppstar+\pploop)\mat{O}(\pstar+\ploop)=\\\ppstar\mat{O}\pstar+\ppstar\mat{O}\ploop+\pploop\mat{O}\pstar+
    \pploop\mat{O}\ploop,
\end{multline}
which allows for the study of the low-frequency behavior of each of the separate terms.
Analysis of the frequency behavior of the first three terms is quite straightforward and yields
\begin{subequations}
    \begin{eqnarray}
        \ppstar \mat{O} \pstar & = & \order{\alpha^2} \quad k\to 0\,,\\
        \ppstar \mat{O} \ploop & = & \order{1} \quad k\to 0\,,\\
        \pploop \mat{O} \pstar & = & \order{1} \quad k\to 0\,.
    \end{eqnarray}
\end{subequations}
Analysis of the last term in \eqref{eq:decomposed_O} requires special care.
It is known that when decomposing $\mat{K}_k$ as
\begin{equation}
    \mat{K}_k = \mat{K}_0 + \mat{K}^\prime_k\,,
\end{equation}
where $\mat{K}_0$ is the static limit of $\mat{K}_k$ and $\mat{K}^\prime_k =
\mat{K}_{k}-\mat{K}_{0}$ is the dynamic remainder,
$\mat{K}^\prime_k=\order{k^2}$ as $k\to 0$
\cite{bogaertLowFrequencyScaling2011}. When using this decomposition in
\eqref{eq:discretizedmfie}, it can be verified that $\mat{K}_{0}$ satisfies
\begin{equation}\label{eq:fundamental}
    \pploop \left( \frac{ \mat{G}^\T}{2} - \mat{K}_{0} \right) \left( \mat{G}^\T \right)^{-1} \left( \frac{ \mat{G}^\T }{2} + \mat{K}_{0} \right) \ploop=\matO\,.
\end{equation}
The above equation holds the key to unlocking a frequency-stable \ac{MFIE}.
The proof of property \eqref{eq:fundamental}
is provided in Appendix~\ref{app:proof_cancellation}.
The term $\pploop\mat{O}\ploop$ can now be studied. To this end, note that
\begin{eqnarray}
 &   &   \alpha^2 \pploop\mat{O}\ploop= \nonumber \\
 & = &   \pploop \left( \frac{ \mat{G}^\T }{2} - \mat{K}_{0} \right) \left( \mat{G}^\T \right)^{-1} \left( \frac{ \mat{G}^\T }{2} + \mat{K}_{0} \right) \ploop \nonumber \\
 &   & + \pploop \left( \frac{ \mat{G}^\T }{2} - \mat{K}_{0} \right) \left( \mat{G}^\T \right)^{-1} \left( \mat{K}^\prime_k \right) \ploop \nonumber  \\
 &   & - \pploop \left( \mat{K}^\prime_{-\jm k} \right) \left( \mat{G}^\T \right)^{-1} \left( \frac{ \mat{G}^\T }{2} + \mat{K}_{0} \right) \ploop  \\
 &   & - \pploop \left( \mat{K}^\prime_{-\jm k} \right) \left( \mat{G}^\T \right)^{-1} \left(  \mat{K}^\prime_k \right) \ploop \nonumber\\
 & = & 0 + \order{k^2} + \order{k^2}  - \order{k^4}, \nonumber
\end{eqnarray}
which completes the low-frequency analysis of the overall operator
\begin{align}
    \mat{O} &= \ppstar\mat{O}\pstar+\ppstar\mat{O}\ploop+\pploop\mat{O}\pstar+\pploop\mat{O}\ploop \nonumber\\
            &= \order{\alpha^2}+\order{1}+\order{1}+\order{\frac{k^2}{\alpha^2}}\,.\label{eq:condanalysis}
\end{align}
To choose $\alpha$, in addition to the conditioning constraint imposed
by \eqref{eq:condanalysis}, we need to consider the physical scaling of the
current, which for a plane wave excitation, is
\cite{qianEnhancedAEFIEPerturbation2010}
\begin{eqnarray}
    \ploop \Jvec & = & \order{1}\,,\\
    \pstar \Jvec & = & \order{k}\,.
\end{eqnarray}
These scaling laws reveal that for a
standard formulation, a severe numerical cancellation is expected due to the
fact that the non-solenoidal component of the current (which scales as
$\order{k}$) will disappear when stored alongside the solenoidal component (which scales as
$\order{1}$). Instead, for the regularized formulation proposed here, the equation
is solved for $\vec{i} = \mat{M}^{-1} \Jvec$, which scales as
\begin{eqnarray}
    \ploop \vec{i} & = & \order{\alpha}\,,\\
    \pstar \vec{i} & = & \order{k / \alpha}\,.
\end{eqnarray}
It is now evident that by setting $\alpha = \sqrt{k}$, the above scaling behaviors become
\begin{eqnarray}
    \ploop \vec{i} & = & \order{\sqrt{k}} \label{eq:planewave_RHS_loop}\,,\\
    \pstar \vec{i} & = & \order{\sqrt{k}} \label{eq:planewave_RHS_star}\,,
\end{eqnarray}
eliminating the low frequency cancellation and, at the same time, stabilizing the matrix at low frequencies.
The latter is seen upon inserting the new scalings into \eqref{eq:condanalysis}:
\begin{align}
    \mat{O} &= \order{\alpha^2}+\order{1}+\order{1}+\order{\frac{k^2}{\alpha^2}} \nonumber\\
            &= \order{k}+\order{1}+\order{1}+\order{k}\,.
\end{align}
The deficiency of the MFIE in the static regime also is solved by the scheme proposed here. In fact, using \eqref{eq:condanalysis} when $k=0$ we obtain
\begin{equation}
    \mat{O}\ploop=\ppstar\mat{O}\ploop\,,
\end{equation}
which proves the existence of an exact matrix null-space in statics of dimension exactly equal to that of the harmonic subspace.

Summarizing, the proposed \ac{MFIE} resolves the three main issues of prior standard and non-standard \ac{MFIE} formulations and now can be linearly combined with \acp{EFIE} using projectors.

\section{A new \ac{CFIE}}\label{sec:cfie}

The theoretical developments of the previous sections resulted in a magnetic field
operator that can be stably discretized for arbitrarily low frequencies using standard
integration rules. The electric counterpart of this operator
was obtained in \cite{andriulliWellConditionedElectricField2013}. We will now combine
these two operators, first proving the resonance-free
property of their continuous combination at high frequencies, and then showing their
compatibility at arbitrarily low frequencies.

Standard \cald \ac{CFIE} equations use a localization strategy
for the \ac{EFIE} component to obtain a resonance-free
equation \cite{adamsPhysicalAnalyticalProperties2004,contopanagosWellconditionedBoundaryIntegral2002}. Here, we follow the Yukawa-\cald approach in
\cite{contopanagosWellconditionedBoundaryIntegral2002}. When the Yukawa-\cald \ac{EFIE} is linearly combined with the new magnetic
operator defined in \Cref{sec:mfie}, the following symmetric
Yukawa-\cald \ac{CFIE} is obtained:
\begin{multline}
 \left( \eta^2\left(\frac{\Iop}{2}-\Kop_{-\jm k}\right) \left(\frac{\Iop}{2}+\Kop_{k}\right)(k)+\Top_{-\jm k}\Top_{k}\right)(\Jv)=\\
  \left(\frac{\Iop}{2}-\Kop_{-\jm k}\right)\left(\n\times\Hv^i\right)+\Top_{-\jm k}\left(\n\times\Ev^i\right).\label{eq:newCFIE}
\end{multline}
To demonstrate that this equation represents a valid \cald{CFIE}, i.e. is free from internal resonances, we prove in Appendix~\ref{sec:resfree} that the operator
\begin{equation}\label{eq:cfieyukOP}
    \left(\eta^2\left(\frac{\Iop}{2}-\Kop_{-\jm k}\right) \left(\frac{\Iop}{2}+\Kop_k\right)(k)+\Top_{-\jm k}\Top_k\right)
\end{equation}
can be inverted for any $k$.


The discretization of the proposed Yukawa-\cald \ac{CFIE} follows directly from that of the new \ac{MFIE} in \Cref{sec:mfie} and
that of the \ac{EFIE} in \cite{andriulliWellConditionedElectricField2013}:
\begin{eqnarray}
& & \eta^2 \matdual{M}^\T \left( \frac{ \mat{G}^\T }{2} - \mat{K}_{-\jm k} \right) \left( \mat{G}^\T \right)^{-1} \left( \frac{ \mat{G}^\T }{2} + \mat{K}_{k} \right) \mat{M} \vec{i} \nonumber \\
& & + \matdual{M}^\T \matdual{T} \matdual{M} \left( \mat{G} \right)^{-1} \mat{M}^\T \mat{T} \mat{M} \vec{i} \nonumber \\
& = & \eta^2 \matdual{M}^\T \left( \frac{ \mat{G}^\T }{2} - \mat{K}_{-\jm k} \right) \left( \mat{G}^\T \right)^{-1} \vec{v}_h \nonumber \\
& & + \matdual{M}^\T \matdual{T} \matdual{M} \left( \mat{G} \right)^{-1} \mat{M}^\T \vec{v}_e\,. \label{eq:discretizedcfie}
\end{eqnarray}
Here $\alpha=1$ and $\alpha=\sqrt{k}$ in the high and low
frequency regime, respectively. We next study the latter more in detail.


Scaling in the latter regime follows from the results of the previous section:
\begin{equation}
\begin{split}
 &\eta^2 \matdual{M}^\T \left( \frac{ \mat{G}^\T }{2} - \mat{K}_{-\jm k} \right) \left( \mat{G}^\T \right)^{-1} \left( \frac{ \mat{G}^\T }{2} + \mat{K}_{k} \right) \mat{M} \vec{i}\\
& \quad  + \matdual{M}^\T \matdual{T} \matdual{M} \left( \mat{G} \right)^{-1} \mat{M}^\T \mat{T} \mat{M} \vec{i} \\
 =&-\jm \left(\matdual{P}^{\Sigma H}\matdual{T}_s\matdual{P}^{\Sigma H}\right)\mat{G}^{-1}\mat{T}_h + \jm
  \matdual{T}_h\mat{G}^{-1}\left(\mat{P}^{\Lambda H}\mat{T}_s\mat{P}^{\Lambda H}\right) + \\
& \quad  \jm \left(\matdual{P}^{\Sigma H}\matdual{T}_s\matdual{P}^{\Sigma H}\right)\mat{G}^{-1}\left(\mat{P}^{\Lambda H}\mat{T}_s\mat{P}^{\Lambda H}\right) +\\
& \quad \eta^2\matdual{P}^{\Sigma H} \left( \frac{ \mat{G}^\T }{2} - \mat{K}_{0} \right) \left( \mat{G}^\T \right)^{-1} \left( \frac{ \mat{G}^\T }{2} + \mat{K}_{0} \right) \jm \mat{P}^{\Sigma} + \\
& \quad  \eta^2 \jm \matdual{P}^{\Lambda} \left( \frac{ \mat{G}^\T }{2} - \mat{K}_{0} \right) \left( \mat{G}^\T \right)^{-1} \left( \frac{ \mat{G}^\T }{2} + \mat{K}_{0} \right)\mat{P}^{\Lambda H} +\\
&\quad O(k)\\
=& \order{1}+\order{1}+\order{1}+\order{1}+\order{1}+\order{k}\,.
\end{split}
\end{equation}
Combining this result with the corresponding \acl{RHS} scalings \eqref{eq:planewave_RHS_loop} and \eqref{eq:planewave_RHS_star} proves the overall low-frequency stability of new \ac{CFIE}.

\section{Numerical Results}\label{sec:numres}

This section presents numerical results that validate the above properties of the proposed \ac{MFIE} and \ac{CFIE}.

\begin{figure}
    \centering
    \includegraphics[width=\widthscale\columnwidth]{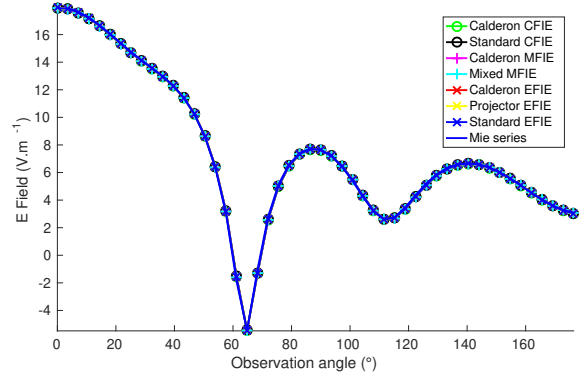}
    \caption{Comparison of the far field scattered by \iac{PEC} sphere of radius \SI{1}{\meter}
    discretized with an average edge size of \SI{0.15}{\meter} and excited by a plane wave
    oscillating at \SI{200}{\mega\hertz}.}
    \label{fig:sphere_far_field_high_freq}
\end{figure}

\begin{figure}
    \centering
    \includegraphics[width=\widthscale\columnwidth]{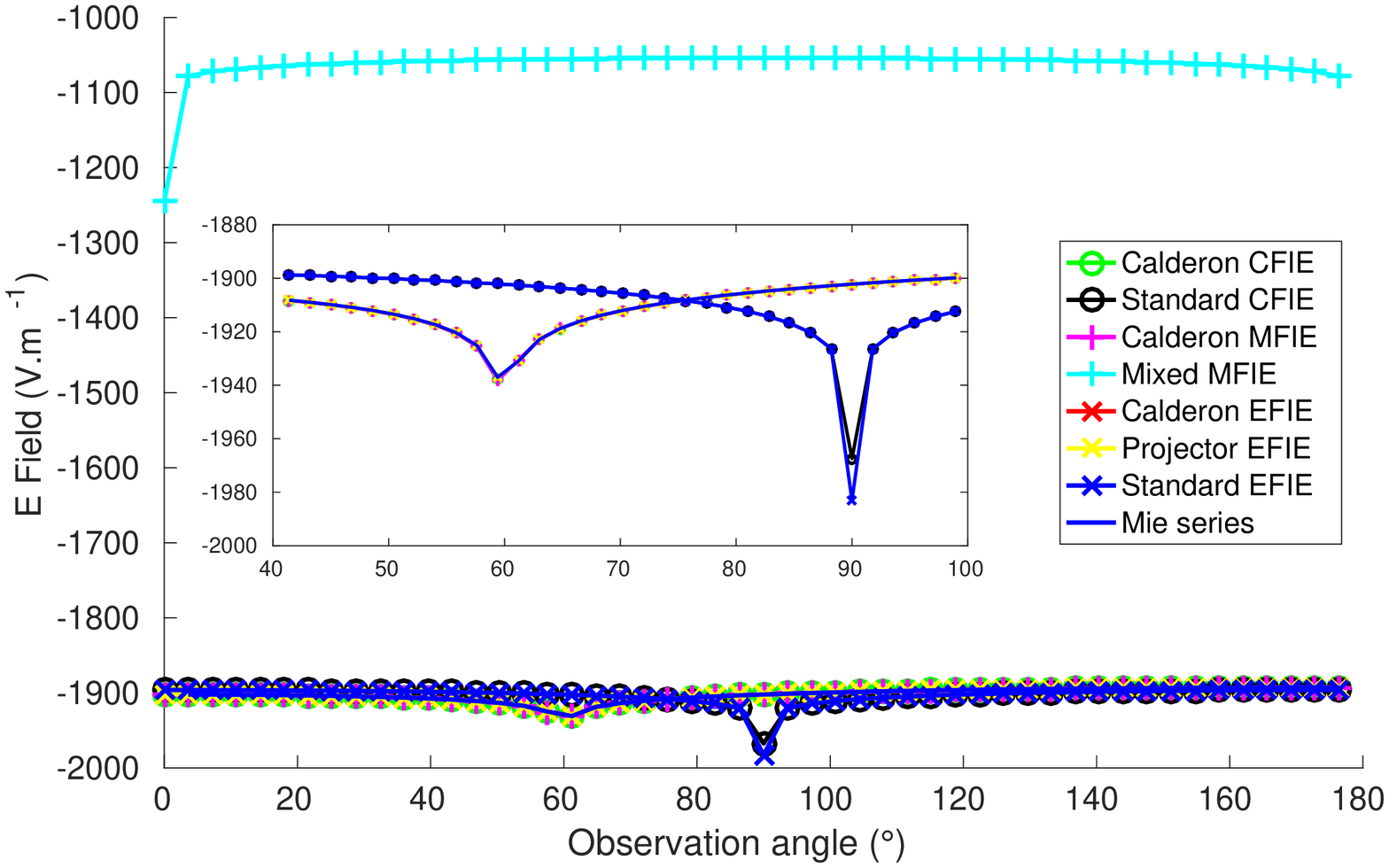}
    \caption{Comparison of the far field scattered by \iac{PEC} sphere of radius \SI{1}{\meter}
    discretized with an average edge size of \SI{0.15}{\meter} and excited by a plane wave
    oscillating at \SI{1e-40}{\hertz}.}
    \label{fig:sphere_far_field_low_freq}
\end{figure}

The first set of tests involve \iac{PEC} sphere of radius $\SI{1}{\meter}$. \Cref{fig:sphere_far_field_high_freq} shows
the scattered far field at $f=\SI{200}{\mega\hertz}$ obtained using the new MFIE and CFIE as well as other established formulations (standard \ac{EFIE}, \ac{EFIE} with projectors, \cald \ac{EFIE}
with projectors, Mixed \ac{MFIE}, \ac{CFIE}).
For this high frequency case all formulations deliver accurate
results, thus validating our implementations. A first difference in
performance between our new formulations and their standard counterparts is noted when lowering the frequency.
\Cref{fig:sphere_far_field_low_freq} shows data similar to \Cref{fig:sphere_far_field_high_freq} but for $f = \SI{1e-40}{\hertz}$. It is clear that accuracy
breakdowns occur for the non-projected methods -- the mixed \ac{MFIE}, the \ac{EFIE}, and the
\ac{CFIE} (for the latter two the lack of accuracy also is due to conditioning problems). On the other hand, all projected formulations, including the two new
ones, deliver accurate results for arbitrarily low frequencies.

\begin{figure}
    \centering
    \includegraphics[width=\widthscale\columnwidth]{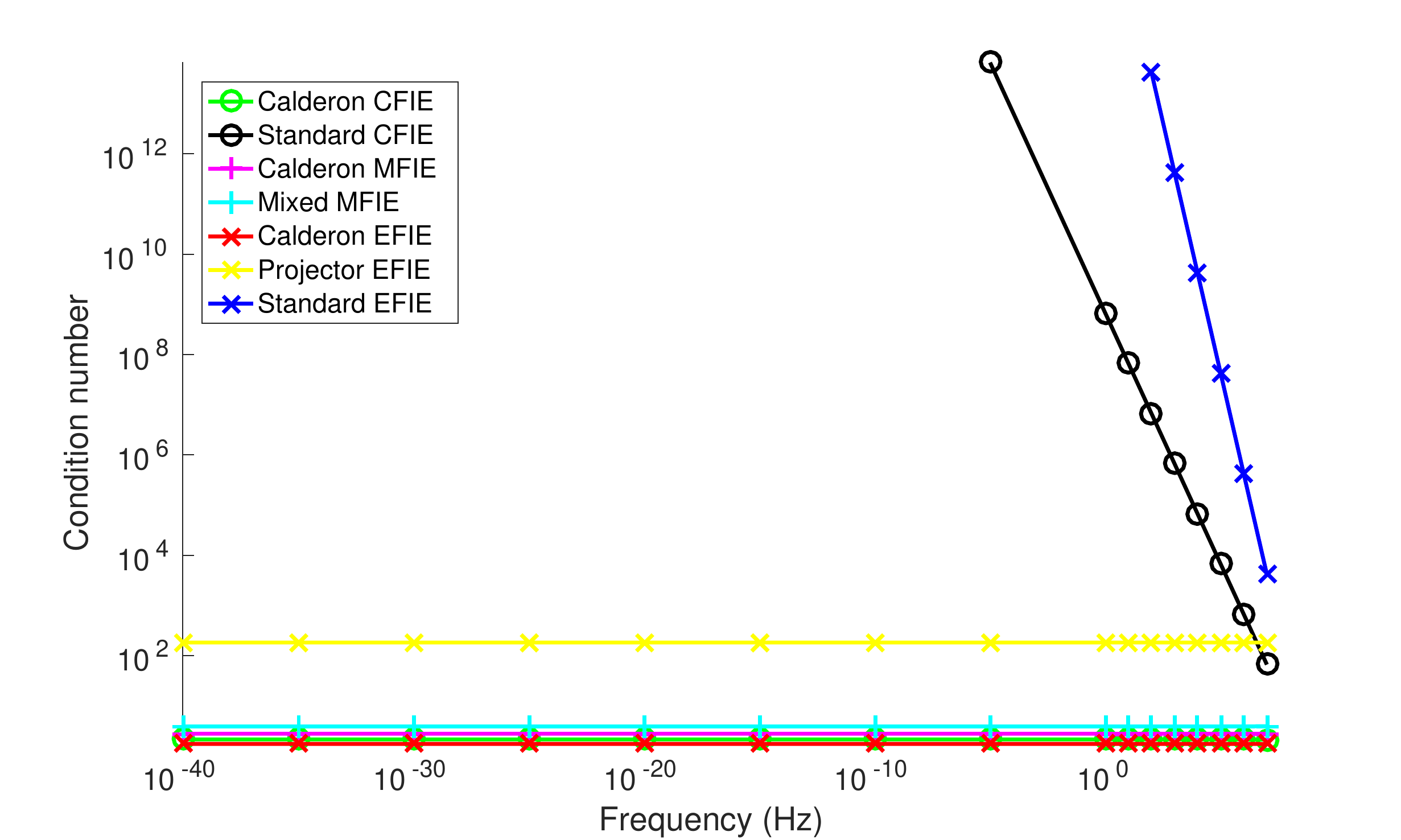}
    \caption{Low frequency behaviour of the conditioning of the different operators on \iac{PEC} sphere of radius \SI{1}{\meter}.  Because of numerical limitations in the computation of
    very high condition numbers ($> \num{1e16}$) some points have been left out.}
    \label{fig:sphere_cond_vs_low_freq}
\end{figure}

The low frequency stability of the new \cald \ac{MFIE} is further demonstrated in
\Cref{fig:sphere_cond_vs_low_freq}, which illustrates the conditioning of the different operators for low frequencies.
It is clear that the
new \ac{MFIE} remains as well-conditioned as its standard counterpart. The
\cald \ac{CFIE} is also low-frequency stable, unlike the standard \ac{CFIE}, which
exhibits a severe ill-conditioning caused by its \ac{EFIE} contribution.

\begin{figure}
    \centering
    \includegraphics[width=\widthscale\columnwidth]{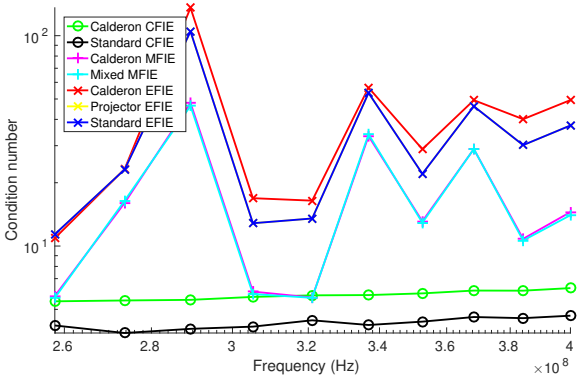}
    \caption{High frequency behaviour of the conditioning of the different operators on
    \iac{PEC} sphere of radius \SI{1}{\meter} sphere illustrating the spurious resonances occurring in non-combined
    formulations. The average edge size of the discretized sphere has been kept
at one-fifth of the wavelength for every simulation.}
    \label{fig:sphere_cond_vs_freq_resonances}
\end{figure}

\Cref{fig:sphere_cond_vs_freq_resonances} shows that, despite its regularized low frequency behavior, the \cald \ac{MFIE} is prone to spurious resonances causing it to become periodically
ill-conditioned. This issue is shared by all non-combined
formulations and can be overcome by combined field strategies.
It is clear from the figure that both the new \cald \ac{CFIE} and
its standard counterpart exhibit resonance-free behaviour.

\begin{figure}
    \centering
    \includegraphics[width=\widthscale\columnwidth]{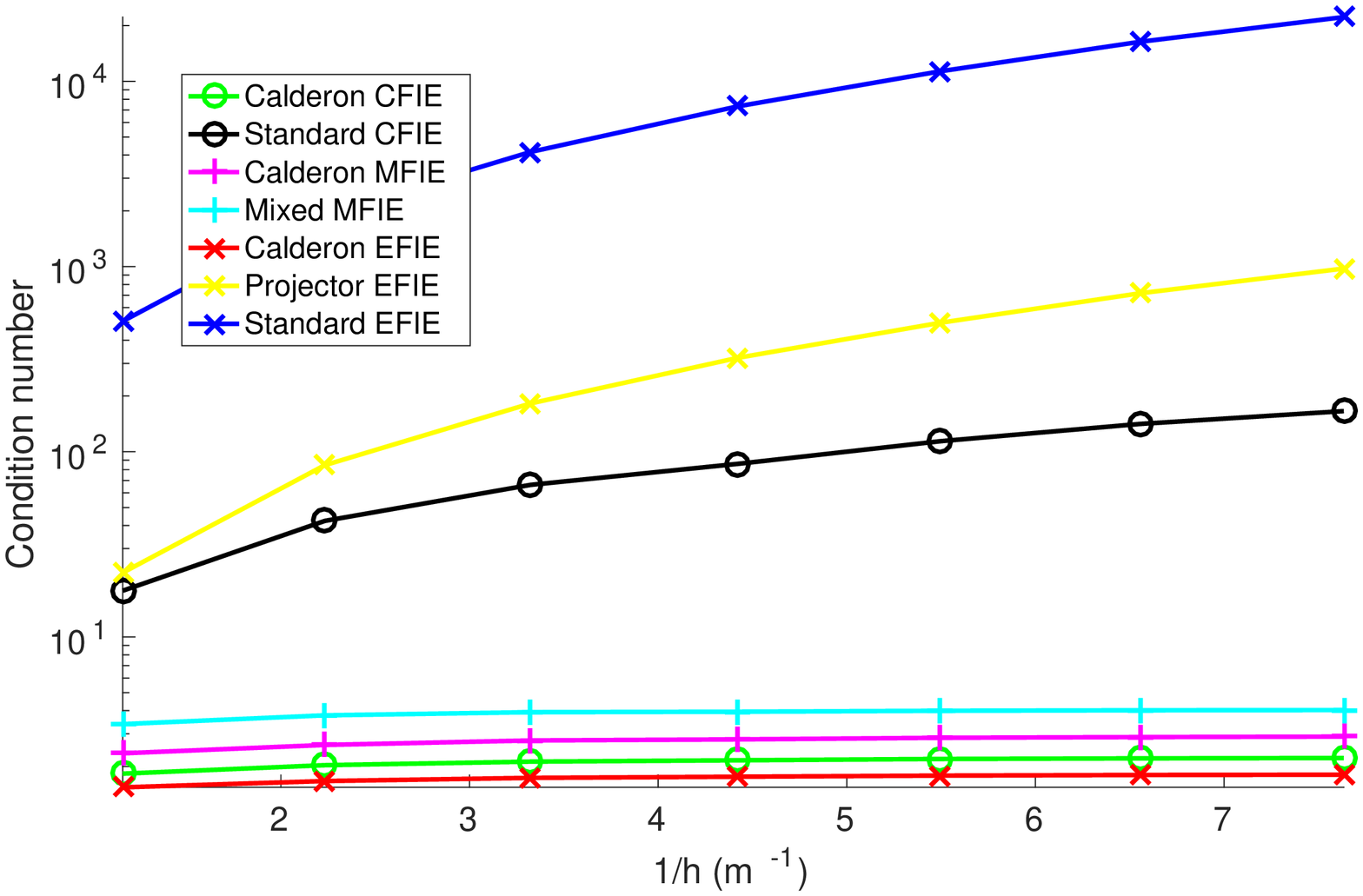}
    \caption{High-refinement behaviour of the conditioning of the different operators on \iac{PEC} sphere of radius \SI{1}{\meter}. The non-resonant frequency has been kept
    constant for all simulations and corresponds to 5 unknowns per wavelength discretization
    for the least refined point.}
    \label{fig:sphere_cond_vs_refinement}
\end{figure}

The last key property to be illustrated is the refinement stability of the
proposed formulations. This property was verified by studying the dependence of the
condition number of the different formulations applied to a unit radius sphere with
increasing discretization density (\Cref{fig:sphere_cond_vs_refinement}).
These results confirm that the
second kind nature of our new formulations renders them immune to the
high-refinement breakdown.

In summary, the above results show that the new \cald \ac{MFIE} yields correct
results for arbitrarily low frequencies and is well conditioned for both low
frequencies and dense discretization. Additionally, when combined with the projector
\cald \ac{EFIE} the new \cald \ac{CFIE}, which is
low frequency stable, immune to dense discretization breakdown, and free from
non-physical resonances, is obtained.

\begin{figure}
    \centering
    \includegraphics[width=\widthscale\columnwidth]{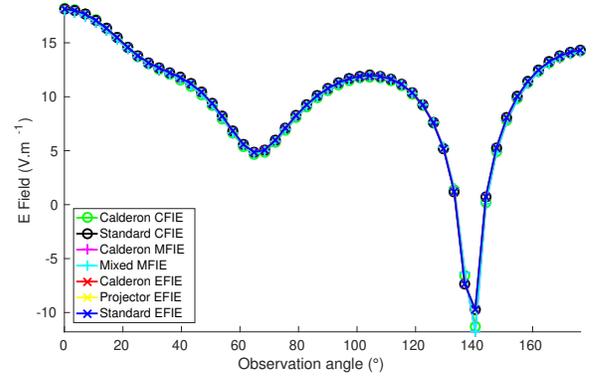}
    \caption{Comparison of the far field scattered by \iac{PEC} square torus with
    an inner radius of \SI{0.5}{\meter} and a tube radius of \SI{0.25}{\meter},
    discretized with an average edge size of \SI{0.15}{\meter} and excited by a plane wave
    oscillating at \SI{200}{\mega\hertz}.}
    \label{fig:torus_far_field_high_freq}
\end{figure}

\begin{figure}
    \centering
    \includegraphics[width=\widthscale\columnwidth]{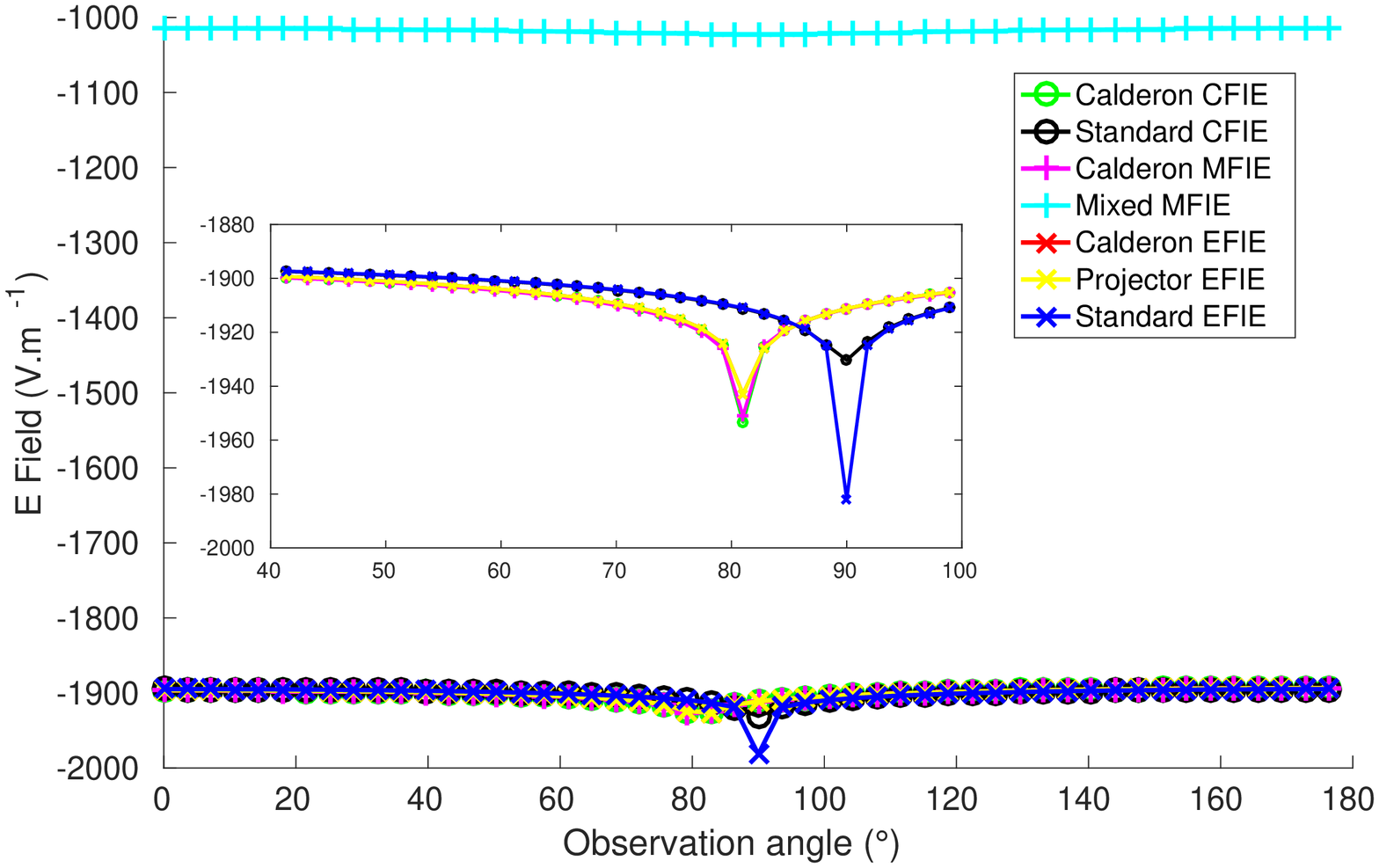}
    \caption{Comparison of the far field scattered by a \ac{PEC} square torus with
    an inner radius of \SI{0.5}{\meter} and a tube radius of \SI{0.25}{\meter},
    discretized  with an average edge size of \SI{0.15}{\meter} and excited by a plane wave
    oscillating at \SI{1e-40}{\hertz}.}
    \label{fig:torus_far_field_low_freq}
\end{figure}

To ensure that the properties illustrated so far still persist for multiply connected structures, many of the previous analyses were repeated for a square torus. The correctness of the formulation has been verified by
studying the far field scattered by the torus at high and very low frequencies,
respectively (\Cref{fig:torus_far_field_high_freq,fig:torus_far_field_low_freq}). Since no analytic solution is
readily available for the square torus, the solution of the \cald \ac{EFIE} was used as a reference
and particular care was taken to avoid frequencies corresponding to an internal resonance.
While the results are similar to those of the sphere, the reader
should be aware that, because of its toroidal and poloidal null-spaces, the
\cald \ac{MFIE} required the usage of a pseudo inversion to obtain current
solutions at very low frequencies.

\begin{figure}
    \centering
    \includegraphics[width=\widthscale\columnwidth]{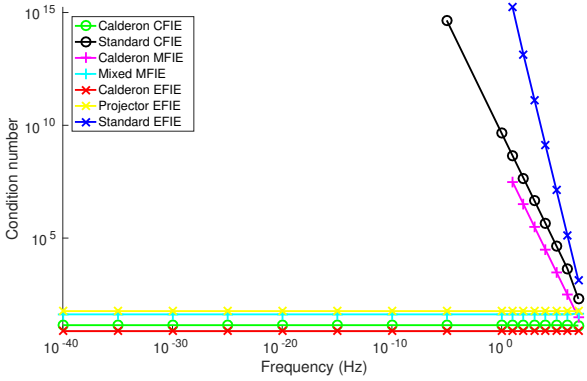}
    \caption{Low frequency behaviour of the conditioning of the different operators on a
    \ac{PEC} square torus with an inner radius of \SI{0.5}{\meter}, a tube
    radius of \SI{0.25}{\meter} and meshed with an average edge length of
    $\SI{0.6}{\meter}$. Because of numerical limitations in the computation
    of very high condition numbers ($> \num{1e16}$) some points have been
    left out.}
    \label{fig:torus_cond_vs_low_freq}
\end{figure}

\begin{figure}
    \centering
    \includegraphics[width=\widthscale\columnwidth]{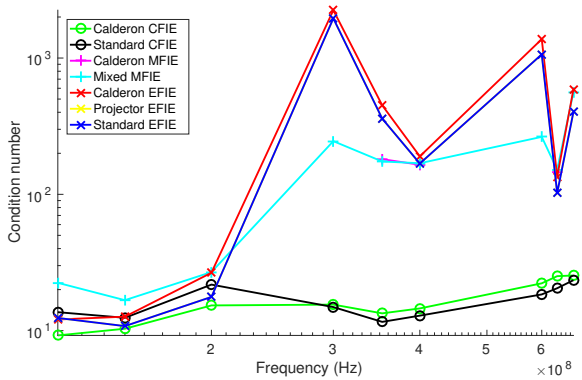}
    \caption{High frequency behaviour of the conditioning of the different operators on a
    \ac{PEC} square torus of inner radius of \SI{0.5}{\meter} and tube radius of
    \SI{0.25}{\meter}, illustrating the resonances of non-combined formulations.
    The average edge size of the discretization has been kept at one-fifth of the
    wavelength for every simulation.}
    \label{fig:torus_cond_vs_freq_resonances}
\end{figure}

\begin{figure}
    \centering
    \includegraphics[width=\widthscale\columnwidth]{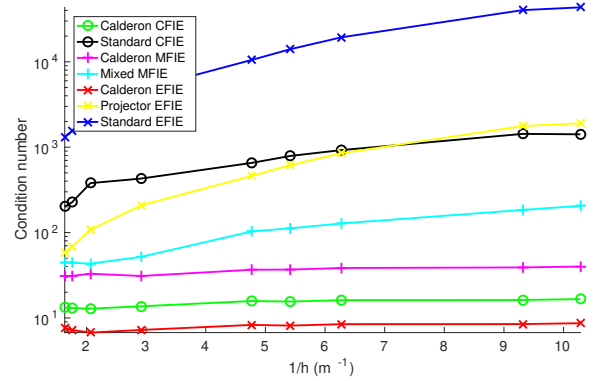}
    \caption{High-refinement behaviour of the conditioning of the different operators on a \ac{PEC}
    square torus of inner radius of \SI{0.5}{\meter} and tube radius of
    \SI{0.25}{\meter}. The non-resonant frequency has been kept
    constant for all simulations and corresponds to a 5 unknowns per wavelength discretization for the least refined point.}
    \label{fig:torus_cond_vs_refinement}
\end{figure}

The low frequency stability of the \cald \ac{MFIE} and \cald \ac{CFIE} on the toroidal
structure are demonstrated in \Cref{fig:torus_cond_vs_low_freq}, while their
resonance free behaviors are illustrated in
\Cref{fig:torus_cond_vs_freq_resonances}. Finally, the resilience of both
formulations to dense discretization breakdown is illustrated in
\Cref{fig:torus_cond_vs_refinement}, which presents the condition number
of the integral operators with increasing discretization of the square
torus.

\begin{figure}
    \centering
    \includegraphics[width=\widthscale\columnwidth]{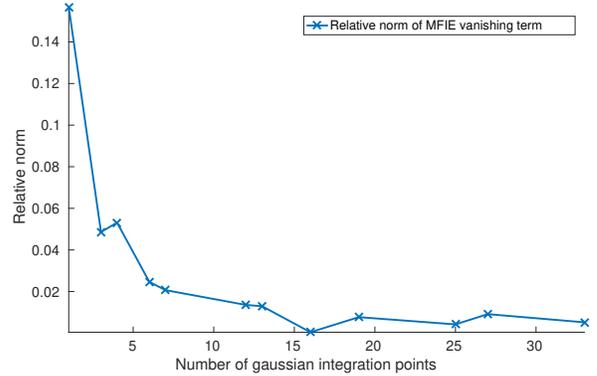}
    \caption{Decay of the relative (with regards to the full operator) norm of
    cancelled out term \eqref{eq:vanishingterm} of the \cald \ac{CFIE} as a
    function of the number of Gaussian integration points. These results correspond
    to a square torus of inner radius \SI{0.5}{\meter} and tube radius
    \SI{0.25}{\meter} simulated at \SI{1e-10}{\hertz}.}
    \label{fig:norm_vs_gaussian_torus}
\end{figure}

\begin{figure}
    \centering
    \includegraphics[width=\widthscale\columnwidth]{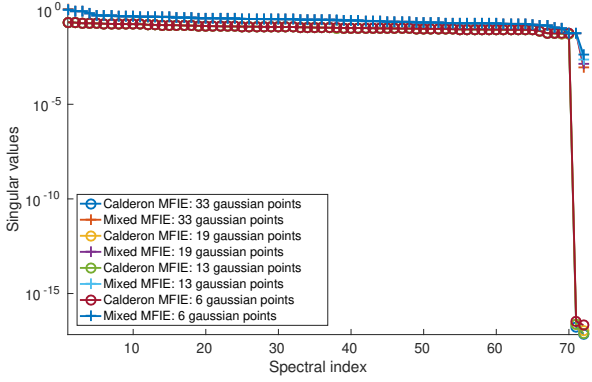}
    \caption{Accuracy of the toroidal and poloidal nullspaces obtained by the
    \cald and Mixed \ac{MFIE} as a function of the number of Gaussian integration points. The results correspond to a
    square torus of inner radius \SI{0.5}{\meter} and tube radius
    \SI{0.25}{\meter} simulated at \SI{1e-10}{\hertz}.}
    \label{fig:svd_vs_gaussian_torus}
\end{figure}

One of the key advantages of the new \cald \ac{MFIE} scheme is that it does not require
extremely accurate numerical integration rules because it allows explicit
cancellation of near-zero terms that are challenging to obtain numerically. The
slow convergence of the standard numerical integration schemes can be seen in
\Cref{fig:norm_vs_gaussian_torus}, in which the ratio
of the norm of the term in \eqref{eq:vanishingterm} to the norm of the full
operator with increasing number of integration points is presented. While this ratio does
decrease with the number of Gaussian quadrature points, it does so very slowly and remains
far from a machine-precision zero value. The effect of these numerical
inaccuracies is evident when comparing the singular value decompositions of the
Mixed \ac{MFIE} and of the new \cald \ac{MFIE} in \Cref{fig:svd_vs_gaussian_torus}. It
is clear that the null singular values corresponding to the toroidal and
poloidal subspaces of the square torus immediately reach the machine precision
zero in the case of the \cald \ac{MFIE}, while for the Mixed \ac{MFIE} they will require
an unreasonably complex integration rules to even remotely resemble a
nullspace.

\begin{figure}
    \centering
    \includegraphics[width=\widthscale\columnwidth]{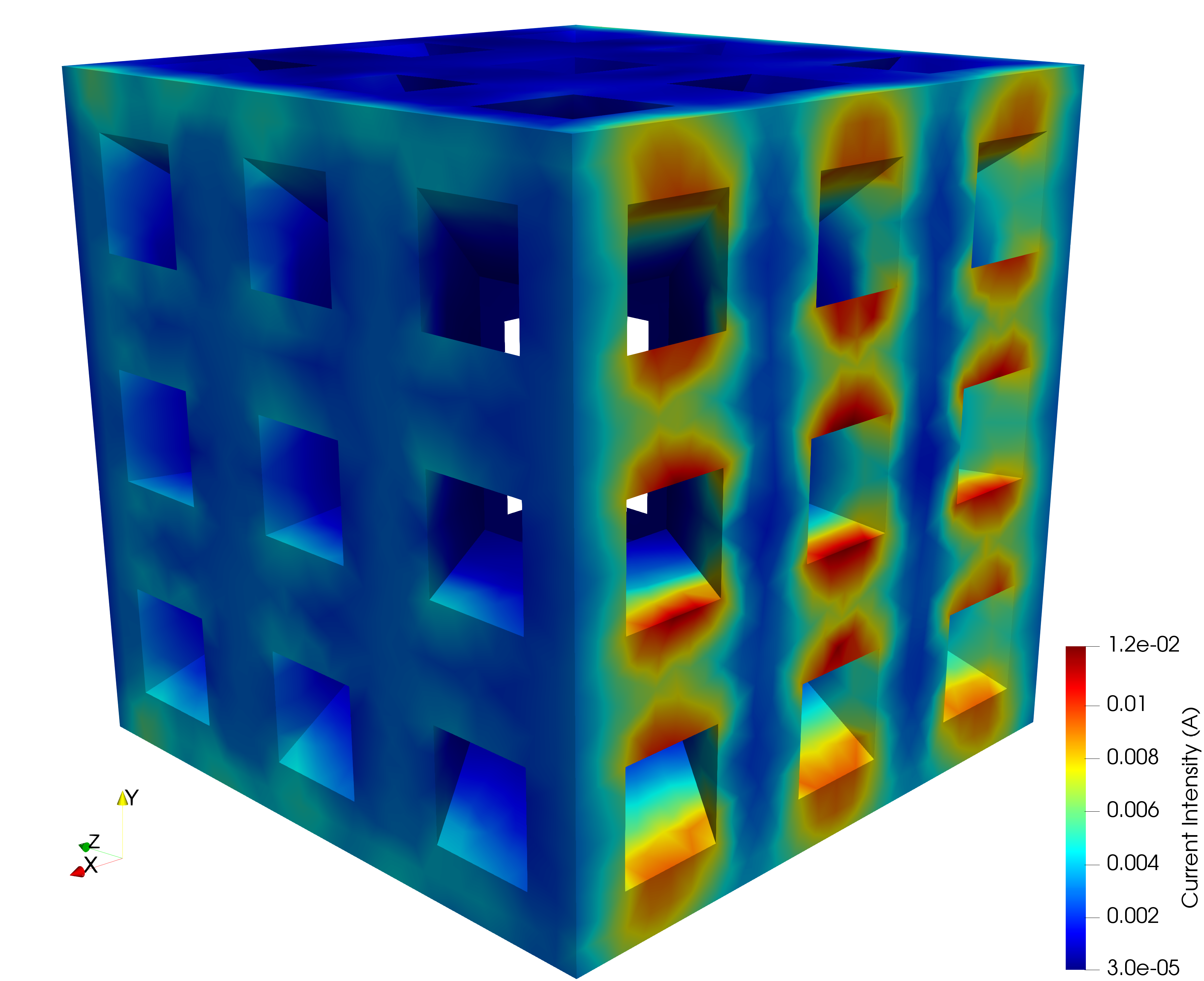}
    \caption{Complex multiply-connected geometry discretized with an average edge
        length of \SI{0.35}{\meter}. The values represented on the geometry
        correspond to the intensity of the current induced on the \ac{PEC} structure by a plane
        wave. The simulating frequency corresponds to $10$ unknowns per wavelength.}
    \label{fig:tesseract}
\end{figure}

\begin{figure}
    \centering
    \includegraphics[width=\widthscale\columnwidth]{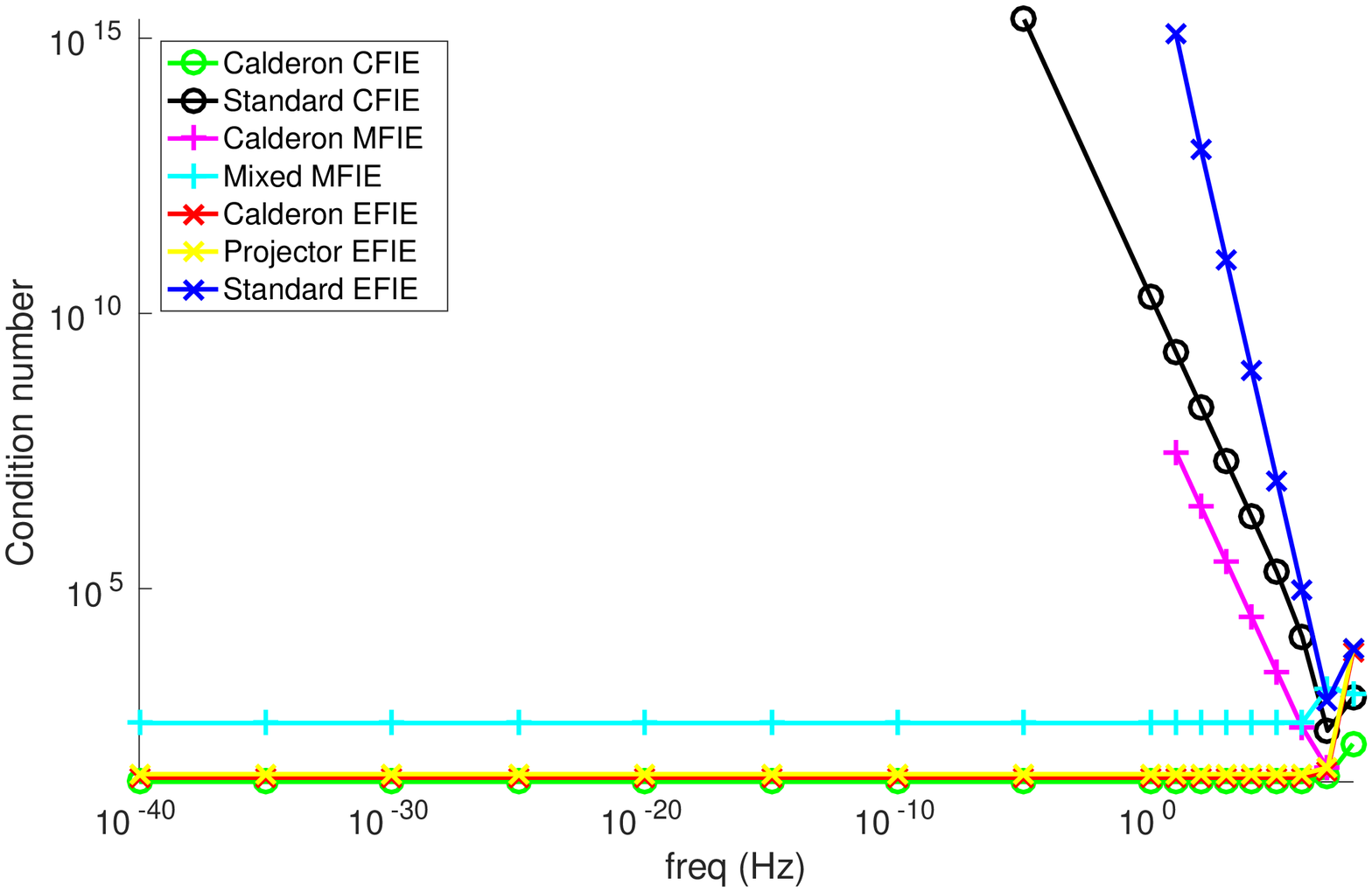}
    \caption{Low frequency behaviour of the conditioning of the different operators on the structure illustrated in \Cref{fig:tesseract}.
        Because of numerical limitations in the computation of very high
        condition numbers ($> \num{1e16}$) some points have been left out.}
    \label{fig:cond_vs_low_freq_tesseract}
\end{figure}

Finally, to demonstrate that our schemes can be readily applied to
more complex problems we studied the low frequency conditioning of our
operators (\Cref{fig:cond_vs_low_freq_tesseract}) for the
complex, multiply connected geometry in \Cref{fig:tesseract}.

\section{Conclusion}\label{sec:conclusion}
This paper presented a new symmetrized MFIE that can be stably and effectively discretized using quasi-Helmholtz projectors.
When linearly combined with a quasi-Helmholtz projector-based \cald \ac{EFIE}, a new \ac{CFIE} that is immune from all drawbacks that plague the majority of existing formulations is obtained. In fact, the proposed CFIE remains well-conditioned both at low frequencies and for  high discretization densities, allows for an accurate solution at extremely low frequencies without requiring special numerical quadrature methods, does not require the detection of global loops when applied to multiply connected geometries, and is provably free from internal resonances. Numerical results confirm the theoretically predicted properties of the proposed equations.

\appendices

\section{Proof of the fundamental matrix relationship \eqref{eq:fundamental}} \label{app:proof_cancellation}

To prove the validity of \eqref{eq:fundamental}, i.e.
\begin{equation}
 \pploop\left( \frac{ \mat{G}^\T }{2} - \mat{K}_{0} \right) \left( \mat{G}^\T \right)^{-1} \left( \frac{ \mat{G}^\T }{2} + \mat{K}_{0} \right)\ploop = \matO\,, \label{eq:vanishingterm}
\end{equation}
we introduce $\pprimal^{\text{Pol}}$, $\pprimal^{\text{Tor}}$, $\pdual^{\text{Pol}}$, $\pdual^{\text{Tor}}$ the orthogonal projectors into the right and left null-spaces of the internal and external \ac{MFIE} operators, i.e.
\begin{align}
    \left( \frac{ \mat{G}^\T }{2} + \mat{K}_{0} \right)\pprimal^{\text{Pol}} &= \matO\,,\\
    \left( \frac{ \mat{G}^\T }{2} - \mat{K}_{0} \right)\pprimal^{\text{Tor}} &= \matO\,,\\
      \pdual^{\text{Pol}}\left( \frac{ \mat{G}^\T }{2} - \mat{K}_{0} \right) &= \matO\,,\\
      \pdual^{\text{Tor}}\left( \frac{ \mat{G}^\T }{2} + \mat{K}_{0} \right) &= \matO\,.
\end{align}
Note that
\begin{align}
    \left( \frac{ \mat{G}^\T }{2} + \mat{K}_{0} \right)\pprimal^{\text{Tor}} &= \mat{G}^\T \pprimal^{\text{Tor}}\,,\\
    \left( \frac{ \mat{G}^\T }{2} - \mat{K}_{0} \right)\pprimal^{\text{Pol}} &= \mat{G}^\T \pprimal^{\text{Pol}}\,,\\
    \pdual^{\text{Tor}}\left( \frac{ \mat{G}^\T }{2} - \mat{K}_{0} \right)   &= \pdual^{\text{Tor}}\mat{G}^\T\,,\\
    \pdual^{\text{Pol}}\left( \frac{ \mat{G}^\T }{2} + \mat{K}_{0} \right)   &= \pdual^{\text{Pol}}\mat{G}^\T\,.
\end{align}
We can then define
\begin{equation}
    \qdual^{\matL}=\pprimal^{\matL \mat H}-\pprimal^{\text{Pol}}-\pprimal^{\text{Tor}}\,,
\end{equation}
which clearly satisfies
\begin{equation}
    \pdual^{\matL}\qdual^{\matL}=\qdual^{\matL}\,,
\end{equation}
since the union of the right null spaces of the internal and external \ac{MFIE}
operators contains all the non-trivial cycles of the structure
\cite{coolsNullspacesMFIECalderon2009}.
Dually,
\begin{equation}
    \qprimal^{\matS}=\pdual^{\matS \mat H}-\pdual^{\text{Pol}}-\pdual^{\text{Tor}}\,,
\end{equation}
satisfies
\begin{equation}
    \pprimal^{\matS}\qprimal^{\matS}=\qprimal^{\matS}\,.
\end{equation}
It follows that
\begin{align}
   &\left( \frac{ \mat{G}^\T }{2} - \mat{K}_{0} \right) \left( \mat{G}^\T \right)^{-1} \left( \frac{ \mat{G}^\T }{2} + \mat{K}_{0} \right)\ploop \nonumber \\
  =&\left( \frac{ \mat{G}^\T }{2} - \mat{K}_{0} \right) \left( \mat{G}^\T \right)^{-1} \left( \frac{ \mat{G}^\T }{2} + \mat{K}_{0} \right)\left(\qdual^{\matL}+\pprimal^{\text{Pol}}+\pprimal^{\text{Tor}}\right) \nonumber \\
  =&\left( \frac{ \mat{G}^\T }{2} - \mat{K}_{0} \right) \left( \mat{G}^\T \right)^{-1} \left( \frac{ \mat{G}^\T }{2} + \mat{K}_{0} \right)\qdual^{\matL}  \nonumber \\
  +&\left( \frac{ \mat{G}^\T }{2} - \mat{K}_{0} \right) \left( \mat{G}^\T \right)^{-1} \left( \frac{ \mat{G}^\T }{2} + \mat{K}_{0} \right)\pprimal^{\text{Tor}} \\
  =&\left( \frac{ \mat{G}^\T }{2} - \mat{K}_{0} \right) \left( \mat{G}^\T \right)^{-1} \left( \frac{ \mat{G}^\T }{2} + \mat{K}_{0} \right)\qdual^{\matL} \nonumber \\
  +&\left( \frac{ \mat{G}^\T }{2} - \mat{K}_{0} \right) \pprimal^{\text{Tor}} \nonumber \\
  =&\left( \frac{ \mat{G}^\T }{2} - \mat{K}_{0} \right) \left( \mat{G}^\T \right)^{-1} \left( \frac{ \mat{G}^\T }{2} + \mat{K}_{0} \right)\qdual^{\matL} \nonumber\,,
\end{align}
and similarly that
\begin{align}
     & \pploop\left( \frac{ \mat{G}^\T }{2} - \mat{K}_{0} \right) \left( \mat{G}^\T \right)^{-1} \left( \frac{ \mat{G}^\T }{2} + \mat{K}_{0} \right) \nonumber \\
    =& \qprimal^{\matS}\left( \frac{ \mat{G}^\T }{2} - \mat{K}_{0} \right) \left( \mat{G}^\T \right)^{-1} \left( \frac{ \mat{G}^\T }{2} + \mat{K}_{0} \right)\,.
\end{align}
Combining the above equations it follows that
\begin{multline}
     \pploop\left( \frac{ \mat{G}^\T }{2} - \mat{K}_{0} \right) \left( \mat{G}^\T \right)^{-1} \left( \frac{ \mat{G}^\T }{2} + \mat{K}_{0} \right)\ploop\\
    = \qprimal^{\matS}\left( \frac{ \mat{G}^\T }{2} - \mat{K}_{0} \right) \left( \mat{G}^\T \right)^{-1} \left( \frac{ \mat{G}^\T }{2} + \mat{K}_{0} \right)\qdual^{\matL}\,. \label{eq:devlopment1}
\end{multline}
In the above expression we now insert the identity matrices $\left(\ploop + \pstar \right)$ and
 $\left( \pploop + \ppstar \right)$ obtaining
\begin{multline}
    \eqref{eq:devlopment1} =  \qprimal^{\matS}\left( \frac{ \mat{G}^\T }{2} - \mat{K}_{0} \right)\left(\pprimal^{\matL \mat H} + \pprimal^{\matS} \right) \left( \mat{G}^\T \right)^{-1}\\
    \left( \pdual^{\matS \mat H} + \pdual^{\matL} \right) \left( \frac{ \mat{G}^\T }{2} + \mat{K}_{0} \right) \qdual^{\matL} \,.
\end{multline}
Given that
\begin{equation}
\qprimal^{\matS}\left( \frac{ \mat{G}^\T }{2} - \mat{K}_{0} \right)\ploop=
\qprimal^{\matS}\pstar\left( \frac{ \mat{G}^\T }{2} - \mat{K}_{0} \right)\ploop
=\matO
\end{equation}
and that
\begin{equation}
 \pploop \left( \frac{ \mat{G}^\T }{2} + \mat{K}_{0} \right) \qdual^{\matL}=
  \pploop \left( \frac{ \mat{G}^\T }{2} + \mat{K}_{0} \right)\ppstar \qdual^{\matL}=\matO
\end{equation}
and considering the property
\begin{equation}
\pstar \left( \mat{G}^\T \right)^{-1} \ppstar =\matO\,,
\end{equation}
we obtain that
\begin{align}
    \eqref{eq:devlopment1} =& \pstar\left( \frac{ \mat{G}^\T }{2} - \mat{K}_{0} \right)\pstar \left( \mat{G}^\T \right)^{-1} \ppstar\left( \frac{ \mat{G}^\T }{2} + \mat{K}_{0} \right)\ppstar \nonumber\\
    =& \matO\,,
\end{align}
which completes the proof.
%

\section{Resonance-free proof for the new \cald \ac{CFIE} operator}\label{sec:resfree}

Since the operator $\left(\frac{\Iop}{2}-\Kop\right)(-\jm k)$ always admits an inverse,  the invertibility of \eqref{eq:cfieyukOP} is equivalent to the invertibility of
\begin{equation}
    \left(\frac{\Iop}{2}+\Kop_k\right)+\left(\frac{\Iop}{2}-\Kop_{-\jm k}\right)^{-1}\Top_{-\jm k}\Top_k.
\end{equation}
Given the anti-commutation property
\begin{equation}
    \Top^{-1}\Kop+\Kop\Top^{-1}=0,
\end{equation}
which follows directly from the second \cald identity $\Top^{-1}\Kop=\Top^{-1}\Kop\Top\Top^{-1}=-\Top^{-1}\Top\Kop\Top^{-1}=-\Kop\Top^{-1}$, and defining
\begin{equation}
    \Aop=\left(\frac{\Iop}{2}-\Kop_{-\jm k}\right)^{-1}\Top_{-\jm k},
\end{equation}
it follows that
\begin{align}
    &\left(\n\times\Aop\right)^{\T}=\left(\n\times\left(\frac{\Iop}{2}-\Kop_{-\jm k}\right)^{-1}\Top_{-\jm k}\right)^{\T} \nonumber \\
    &=\left(\n\times\left(\Top_{-\jm k}^{-1}\left(\frac{\Iop}{2}-\Kop_{-\jm k}\right)\right)^{-1}\right)^{\T} \nonumber \\
    &=\left(\n\times\left(\left(\frac{\Iop}{2}+\Kop_{-\jm k}\right)\Top_{-\jm k}^{-1}\right)^{-1}\right)^{\T} \nonumber \\
    &=\left(\n\times\Top_{-\jm k}\left(\frac{\Iop}{2}+\Kop_{-\jm k}\right)^{-1}\right)^{\T}\\
    &=\left(\left(\frac{\Iop}{2}+\Kop_{-\jm k}\right)^{-1}\right)^{\T}\n\times\Top_{-\jm k} \nonumber \\
    &=-\n\times\left(\frac{\Iop}{2}-\Kop_{-\jm k}\right)^{-1}\n\times\n\times\Top_{-\jm k} \nonumber \\
    &=\n\times\left(\frac{\Iop}{2}-\Kop_{-\jm k}\right)^{-1}\Top_{-\jm k} \nonumber \\
    &=\n\times\Aop. \nonumber
\end{align}
Given this result and the fact that
\begin{equation}
    \n\times\Aop=\n\times\left(\frac{\Iop}{2}-\Kop_{-\jm k}\right)^{-1}\Top_{-\jm k}
\end{equation}
is a real operator, the symmetry implies it being Hermitian, so that
\begin{equation}
    \vec{x}^\dagger\left(\n\times\left(\left(\frac{\Iop}{2}-\Kop_{-\jm k}\right)(\jm k)\right)^{-1}\Top_{-\jm k}\right)\vec{x}
\end{equation}
is real and nonzero.
By leveraging a straightforward extension of Theorem~3.1 in
\cite{brunoElectromagneticIntegralEquations2009}, it follows that
\begin{equation}
    \left(\left(\frac{\Iop}{2}-\Kop_{-\jm k}\right)     \left(\frac{\Iop}{2}+\Kop_{k}\right)+\Top_{-\jm k}\Top_{k}\right)
\end{equation}
is always invertible. Otherwise said, the Yukawa-\cald \ac{CFIE} we propose is resonance free.

\section*{Acknowledgment}

This work has been funded in part by the European Research Council (ERC) under the European Union's Horizon 2020 research and innovation program (ERC project 321, grant No. 724846).

\bibliography{cfie}
\bibliographystyle{IEEEtran}

\end{document}

%% file: macros_compat_paper.tex
\usepackage[OMLmathsfit]{isomath}
\usepackage{bm}
\usepackage{upgreek}
\DeclareMathAlphabet{\mathbbmsl}{U}{bbm}{m}{sl}
\newcommand{\matdual}[1]{\mathbbmsl{#1}}

\newcommand{\mathbcal}[1]{\mathcal{#1}}
\DeclareMathAlphabet{\mathbfsf}{\encodingdefault}{\sfdefault}{bx}{n}

\newcommand{\veg}[1]{\bm{#1}}     
\newcommand{\mat}[1]{\mathsfbfit{#1}} 
\renewcommand{\vec}[1]{\mathsfbfit{#1}} 
\newcommand{\vecop}[1]{\bm{\mathbcal{#1}}} 

\newcommand{\n}{\hat{\veg{n}}}

\newcommand{\vegO}{\veg{0}}
\newcommand{\matO}{\veg{0}}

\newcommand{\dd}{\mathrm{d}}  

\newcommand{\jm}{\mathrm{j}}  

\newcommand{\R}{\mathbb{R}}

\newcommand{\T}{\mathrm{T}}


%% file: own_macros.tex
\newcommand{\matL}{\mat{\Lambda}}   
\newcommand{\matS}{\mat{\Sigma}}

\newcommand{\matId}{\mat{\mathrm{I}}}

\newcommand{\Ev}{\veg e}
\newcommand{\Hv}{\veg h}
\newcommand{\Jv}{\veg j}

\newcommand{\Jvec}{\vec j}

\newcommand{\Top}{\vecop{T}}
\newcommand{\Kop}{\vecop{K}}
\newcommand{\Iop}{\vecop{I}}
\newcommand{\Aop}{\vecop{A}}

\newcommand{\pseudoinv}[1]{#1^{+}}
\newcommand{\transpose}[1]{#1^\T}

\newcommand{\vr}{\veg r}
\newcommand{\vrp}{\veg{r}^\prime}

\newcommand{\cald}{Calder\'on }

\newcommand{\order}[1]{\mathcal{O}(#1)}

\newcommand{\ploop}{\mat{P}^{\matL \mat H}}
\newcommand{\pstar}{\mat{P}^{\matS}}
\newcommand{\ppstar}{\matdual{P}^{\matL}}
\newcommand{\pploop}{\matdual{P}^{\matS \mat H}}

%% file: main.bbl
\begin{thebibliography}{10}
\providecommand{\url}[1]{#1}
\csname url@samestyle\endcsname
\providecommand{\newblock}{\relax}
\providecommand{\bibinfo}[2]{#2}
\providecommand{\BIBentrySTDinterwordspacing}{\spaceskip=0pt\relax}
\providecommand{\BIBentryALTinterwordstretchfactor}{4}
\providecommand{\BIBentryALTinterwordspacing}{\spaceskip=\fontdimen2\font plus
\BIBentryALTinterwordstretchfactor\fontdimen3\font minus
  \fontdimen4\font\relax}
\providecommand{\BIBforeignlanguage}[2]{{%
\expandafter\ifx\csname l@#1\endcsname\relax
\typeout{** WARNING: IEEEtran.bst: No hyphenation pattern has been}%
\typeout{** loaded for the language `#1'. Using the pattern for}%
\typeout{** the default language instead.}%
\else
\language=\csname l@#1\endcsname
\fi
#2}}
\providecommand{\BIBdecl}{\relax}
\BIBdecl

\bibitem{vanbladelElectromagneticFields2007}
J.~G. Van~Bladel, \emph{Electromagnetic Fields}, 2nd~ed., ser. IEEE Press
  series on electromagnetic wave theory.\hskip 1em plus 0.5em minus 0.4em\relax
  {IEEE Press}.

\bibitem{raoElectromagneticScatteringSurfaces1982}
S.~Rao, D.~Wilton, and A.~Glisson, ``Electromagnetic scattering by surfaces of
  arbitrary shape,'' vol.~30, no.~3, pp. 409--418.

\bibitem{coolsAccurateConformingMixed2011}
K.~Cools, F.~P. Andriulli, D.~De~Zutter, and E.~Michielssen, ``Accurate and
  {{Conforming Mixed Discretization}} of the {{MFIE}},'' vol.~10, pp. 528--531.

\bibitem{yunhuazhangMagneticFieldIntegral2003}
{Yunhua Zhang}, {Tie Jun Cui}, {Weng Cho Chew}, and {Jun-Sheng Zhao},
  ``Magnetic field integral equation at very low frequencies,'' vol.~51, no.~8,
  pp. 1864--1871.

\bibitem{qianEnhancedAEFIEPerturbation2010}
Z.-G. Qian and W.~C. Chew, ``Enhanced {{A}}-{{EFIE With Perturbation
  Method}},'' vol.~58, no.~10, pp. 3256--3264.

\bibitem{qianQuantitativeStudyLow2008}
Z.~G. Qian and W.~C. Chew, ``A quantitative study on the low frequency
  breakdown of {{EFIE}},'' vol.~50, no.~5, pp. 1159--1162.

\bibitem{wiltonImprovingElectricField1981}
D.~Wilton and A.~Glisson, ``On improving the electric field integral equation
  at low frequencies,'' vol.~24.

\bibitem{vecchiLoopstarDecompositionBasis1999}
G.~Vecchi, ``Loop-star decomposition of basis functions in the discretization
  of the {{EFIE}},'' vol.~47, no.~2, pp. 339--346.

\bibitem{zhaoIntegralEquationSolution2000}
J.-S. Zhao and W.~C. Chew, ``Integral equation solution of {{Maxwell}}'s
  equations from zero frequency to microwave frequencies,'' vol.~48, no.~10,
  pp. 1635--1645.

\bibitem{leeLoopStarBasis2003}
J.-F. Lee, R.~Lee, and R.~Burkholder, ``Loop star basis functions and a robust
  preconditioner for {{EFIE}} scattering problems,'' vol.~51, no.~8, pp.
  1855--1863.

\bibitem{eibertIterativesolverConvergenceLoopstar2004}
T.~F. Eibert, ``Iterative-solver convergence for loop-star and loop-tree
  decompositions in method-of-moments solutions of the electric-field integral
  equation,'' vol.~46, no.~3, pp. 80--85.

\bibitem{wiltonImprovingStabilityElectric1981}
D.~Wilton and A.~Glisson, ``On improving the stability of the electric field
  integral equation at low frequency,'' in \emph{Proc. {{IEEE Antennas}} and
  {{Propagation Soc}}. {{National Symp}}}, pp. 124--133.

\bibitem{andriulliSolvingEFIELow2010}
F.~P. Andriulli, A.~Tabacco, and G.~Vecchi, ``Solving the {{EFIE}} at {{Low
  Frequencies With}} a {{Conditioning That Grows Only Logarithmically With}}
  the {{Number}} of {{Unknowns}},'' vol.~58, no.~5, pp. 1614--1624.

\bibitem{andriulliMultiplicativeCalderonPreconditioner2008}
F.~P. Andriulli, K.~Cools, H.~Bagci, F.~Olyslager, A.~Buffa, S.~Christiansen,
  and E.~Michielssen, ``A {{Multiplicative Calderon Preconditioner}} for the
  {{Electric Field Integral Equation}},'' vol.~56, no.~8, pp. 2398--2412.

\bibitem{andriulliLoopStarLoopTreeDecompositions2012}
F.~P. Andriulli, ``Loop-{{Star}} and {{Loop}}-{{Tree Decompositions}}:
  {{Analysis}} and {{Efficient Algorithms}},'' vol.~60, no.~5, pp. 2347--2356.

\bibitem{qianAugmentedElectricField2008}
Z.~G. Qian and W.~C. Chew, ``An augmented electric field integral equation for
  high-speed interconnect analysis,'' vol.~50, no.~10, pp. 2658--2662.

\bibitem{zhuRigorousSolutionLowFrequency2011}
J.~Zhu and D.~Jiao, ``A {{Rigorous Solution}} to the {{Low}}-{{Frequency
  Breakdown}} in {{Full}}-{{Wave Finite}}-{{Element}}-{{Based Analysis}} of
  {{General Problems Involving Inhomogeneous Lossless}}/{{Lossy Dielectrics}}
  and {{Nonideal Conductors}},'' vol.~59, no.~12, pp. 3294--3306.

\bibitem{vipianaMultiresolutionSystemRao2007}
F.~Vipiana, G.~Vecchi, and P.~Pirinoli, ``A {{Multiresolution System}} of
  {{Rao}} amp;ndash;{{Wilton}} amp;ndash;{{Glisson Functions}},'' vol.~55,
  no.~3, pp. 924--930.

\bibitem{andriulliHierarchicalBasesNonhierarchic2008}
F.~P. Andriulli, F.~Vipiana, and G.~Vecchi, ``Hierarchical {{Bases}} for
  {{Nonhierarchic}} 3-{{D Triangular Meshes}},'' vol.~56, no.~8, pp.
  2288--2297.

\bibitem{chenMultiresolutionCurvilinearRao2009}
R.~Chen, J.~Ding, D.~Z. Ding, Z.~H. Fan, and D.~Wang, ``A {{Multiresolution
  Curvilinear Rao}}–{{Wilton}}–{{Glisson Basis Function}} for {{Fast
  Analysis}} of {{Electromagnetic Scattering}},'' vol.~57, no.~10, pp.
  3179--3188.

\bibitem{christiansenPreconditionerElectricField2002}
S.~H. Christiansen and J.-C. Nedelec, ``A {{Preconditioner}} for the {{Electric
  Field Integral Equation Based}} on {{Calderon Formulas}},'' vol.~40, no.~3,
  pp. 1100--1135.

\bibitem{contopanagosWellconditionedBoundaryIntegral2002}
H.~Contopanagos, B.~Dembart, M.~Epton, J.~Ottusch, V.~Rokhlin, J.~Visher, and
  S.~Wandzura, ``Well-conditioned boundary integral equations for
  three-dimensional electromagnetic scattering,'' vol.~50, no.~12, pp.
  1824--1830.

\bibitem{adamsPhysicalAnalyticalProperties2004}
R.~Adams, ``Physical and {{Analytical Properties}} of a {{Stabilized Electric
  Field Integral Equation}},'' vol.~52, no.~2, pp. 362--372.

\bibitem{darbasGeneralizedCombinedField2006}
M.~Darbas, ``Generalized combined field integral equations for the iterative
  solution of the three-dimensional {{Maxwell}} equations,'' vol.~19, no.~8,
  pp. 834--839.

\bibitem{stephansonPreconditionedElectricField2009}
M.~B. Stephanson and J.-F. Lee, ``Preconditioned {{Electric Field Integral
  Equation Using Calderon Identities}} and {{Dual Loop}}/{{Star Basis
  Functions}},'' vol.~57, no.~4, pp. 1274--1279.

\bibitem{suyanEFIEAnalysisLowFrequency2010}
{Su Yan}, {Jian-Ming Jin}, and {Zaiping Nie}, ``{{EFIE Analysis}} of
  {{Low}}-{{Frequency Problems With Loop}}-{{Star Decomposition}} and
  {{Calder}}\&\#{{x00F3}};n {{Multiplicative Preconditioner}},'' vol.~58,
  no.~3, pp. 857--867.

\bibitem{5654576}
P.~Yla-Oijala, S.~P. Kiminki, and S.~Jarvenpaa, ``Calderon preconditioned
  surface integral equations for composite objects with junctions,'' \emph{IEEE
  Transactions on Antennas and Propagation}, vol.~59, no.~2, pp. 546--554, Feb
  2011.

\bibitem{DOBBELAERE2015355}
D.~Dobbelaere, D.~D. Zutter, J.~V. Hese, J.~Sercu, T.~Boonen, and H.~Rogier,
  ``A calderón multiplicative preconditioner for the electromagnetic
  poincaré–steklov operator of a heterogeneous domain with scattering
  applications,'' \emph{Journal of Computational Physics}, vol. 303, pp. 355 --
  371, 2015.

\bibitem{6934990}
J.~Markkanen, ``Discrete helmholtz decomposition for electric current volume
  integral equation formulation,'' \emph{IEEE Transactions on Antennas and
  Propagation}, vol.~62, no.~12, pp. 6282--6289, Dec 2014.

\bibitem{epsteinDebyeSourcesNumerical2010}
C.~L. Epstein and L.~Greengard, ``Debye sources and the numerical solution of
  the time harmonic {{Maxwell}} equations,'' vol.~63, no.~4, pp. 413--463.

\bibitem{chewIntegralEquationMethods2008}
W.~C. Chew, M.~S. Tong, and B.~Hu, \emph{Integral {{Equation Methods}} for
  {{Electromagnetic}} and {{Elastic Waves}}}, vol.~3.

\bibitem{bogaertLowFrequencyScalingStandard2014}
I.~Bogaert, K.~Cools, F.~P. Andriulli, and H.~Bagci, ``Low-{{Frequency
  Scaling}} of the {{Standard}} and {{Mixed Magnetic Field}} and muller
  {{Integral Equations}},'' vol.~62, no.~2, pp. 822--831.

\bibitem{vicoDecoupledPotentialIntegral2016}
F.~Vico, M.~Ferrando, L.~Greengard, and Z.~Gimbutas, ``The {{Decoupled
  Potential Integral Equation}} for {{Time}}-{{Harmonic Electromagnetic
  Scattering}},'' vol.~69, no.~4, pp. 771--812.

\bibitem{sunCalderonMultiplicativePreconditioned2013}
S.~Sun, Y.~G. Liu, W.~C. Chew, and Z.~Ma, ``Calderón {{Multiplicative
  Preconditioned EFIE With Perturbation Method}},'' vol.~61, no.~1, pp.
  247--255.

\bibitem{chengAugmentedEFIENormally2015}
J.~Cheng, R.~J. Adams, J.~C. Young, and M.~A. Khayat, ``Augmented {{EFIE With
  Normally Constrained Magnetic Field}} and {{Static Charge Extraction}},''
  vol.~63, no.~11, pp. 4952--4963.

\bibitem{dasModifiedSeparatedPotential2016}
A.~Das and D.~Gope, ``Modified {{Separated Potential Integral Equation}} for
  {{Low}}-{{Frequency EFIE Conditioning}},'' vol.~64, no.~4, pp. 1394--1403.

\bibitem{andriulliWellConditionedElectricField2013}
F.~P. Andriulli, K.~Cools, I.~Bogaert, and E.~Michielssen, ``On a
  {{Well}}-{{Conditioned Electric Field Integral Operator}} for {{Multiply
  Connected Geometries}},'' vol.~61, no.~4, pp. 2077--2087.

\bibitem{begheinDCStableLargeTime2015}
Y.~Beghein, K.~Cools, and F.~P. Andriulli, ``A {{DC Stable}} and
  {{Large}}-{{Time Step Well}}-{{Balanced TD}}-{{EFIE Based}} on
  {{Quasi}}-{{Helmholtz Projectors}},'' vol.~63, no.~7, pp. 3087--3097.

\bibitem{begheinDCStableWellBalancedCalderon2015}
------, ``A {{DC}}-{{Stable}}, {{Well}}-{{Balanced}}, {{Calderón
  Preconditioned Time Domain Electric Field Integral Equation}},'' vol.~63,
  no.~12, pp. 5650--5660.

\bibitem{begheinHandlingLowfrequencyBreakdown2015}
Y.~Beghein, R.~Mitharwal, K.~Cools, and F.~P. Andriulli, ``Handling the
  low-frequency breakdown of the {{PMCHWT}} integral equation with the
  quasi-{{Helmholtz}} projectors,'' in \emph{2015 {{International Conference}}
  on {{Electromagnetics}} in {{Advanced Applications}} ({{ICEAA}})}, pp.
  1534--1537.

\bibitem{begheinRobustLowFrequency2015}
Y.~Beghein, K.~Cools, and F.~P. Andriulli, ``A robust and low frequency stable
  time domain {{PMCHWT}} equation,'' in \emph{2015 {{International Conference}}
  on {{Electromagnetics}} in {{Advanced Applications}} ({{ICEAA}})}, pp.
  954--957.

\bibitem{chewGedankenExperimentsUnderstand2007}
W.~C. Chew and J.~M. Song, ``Gedanken {{Experiments}} to {{Understand}} the
  {{Internal Resonance Problems}} of {{Electromagnetic Scattering}},'' vol.~27,
  no.~8, pp. 457--471.

\bibitem{andriulliMagneticTypeIntegral2014}
F.~P. Andriulli, I.~Bogaert, K.~Cools, and E.~Michielssen, ``A magnetic type
  integral operator which is stable till extremely low frequencies,'' in
  \emph{2014 {{XXXIth URSI General Assembly}} and {{Scientific Symposium}}
  ({{URSI GASS}})}, pp. 1--4.

\bibitem{andriulliWellconditionedCombinedField2013}
------, ``A well-conditioned combined field integral equation based on
  quasi-helmholtz projectors,'' in \emph{2013 {{International Conference}} on
  {{Electromagnetics}} in {{Advanced Applications}} ({{ICEAA}})}, pp. 644--647.

\bibitem{buffaDualFiniteElement2007}
A.~Buffa and S.~Christiansen, ``A dual finite element complex on the
  barycentric refinement,'' vol.~76, no. 260, pp. 1743--1769.

\bibitem{coolsNullspacesMFIECalderon2009}
K.~Cools, F.~Andriulli, F.~Olyslager, and E.~Michielssen, ``Nullspaces of
  {{MFIE}} and {{Calderon Preconditioned EFIE Operators Applied}} to {{Toroidal
  Surfaces}},'' vol.~57, no.~10, pp. 3205--3215.

\bibitem{wiltonTopologicalConsiderationSurface1983}
D.~R. Wilton, ``Topological consideration in surface patch and volume cell
  modeling of electromagnetic scatterers,'' in \emph{Proc. {{URSI Int}}.
  {{Symp}}. {{Electromagn}}. {{Theory}}}, pp. 65--68.

\bibitem{napovAlgebraicMultigridMethod2012}
A.~Napov and Y.~Notay, ``An {{Algebraic Multigrid Method}} with {{Guaranteed
  Convergence Rate}},'' vol.~34, no.~2, pp. A1079--A1109.

\bibitem{coolsImprovingMFIEAccuracy2009}
K.~Cools, F.~P. Andriulli, F.~Olyslager, and E.~Michielssen, ``Improving the
  {{MFIE}}'s accuracy by using a mixed discretization,'' in \emph{2009 {{IEEE
  Antennas}} and {{Propagation Society International Symposium}}}, pp. 1--4.

\bibitem{chenElectromagneticScatteringThreedimensional1990}
Q.~Chen and D.~Wilton, ``Electromagnetic scattering by three-dimensional
  arbitrary complex material/conducting bodies,'' in \emph{Antennas and
  {{Propagation Society International Symposium}}, 1990. {{AP}}-{{S}}.
  {{Merging Technologies}} for the 90's. {{Digest}}.}, pp. 590--593 vol.2.

\bibitem{bogaertLowFrequencyScaling2011}
I.~Bogaert, K.~Cools, F.~P. Andriulli, and D.~De~Zutter, ``Low frequency
  scaling of the mixed {{MFIE}} for scatterers with a non-simply connected
  surface,'' in \emph{Electromagnetics in {{Advanced Applications}}
  ({{ICEAA}}), 2011 {{International Conference}} On}.\hskip 1em plus 0.5em
  minus 0.4em\relax {IEEE}, pp. 951--954.

\bibitem{bogaertLowFrequencyStability2011}
I.~Bogaert, K.~Cools, F.~P. Andriulli, J.~Peeters, and D.~De~Zutter, ``Low
  frequency stability of the mixed discretization of the {{MFIE}},'' in
  \emph{Antennas and {{Propagation}} ({{EUCAP}}), {{Proceedings}} of the 5th
  {{European Conference}} On}.\hskip 1em plus 0.5em minus 0.4em\relax {IEEE},
  pp. 2463--2465.

\bibitem{brunoElectromagneticIntegralEquations2009}
O.~Bruno, T.~Elling, R.~Paffenroth, and C.~Turc, ``Electromagnetic integral
  equations requiring small numbers of {{Krylov}}-subspace iterations,'' vol.
  228, no.~17, pp. 6169--6183.

\end{thebibliography}
